\documentclass[twocolumn]{aastex63}

\accepted{May 8th, 2020}

\submitjournal{ApJ}

\shorttitle{Survey of Extremely-High-Velocity Outflows in SDSS Quasars}
\shortauthors{Rodr\'iguez Hidalgo et al.}

\usepackage[T1]{fontenc}
\usepackage{ae,aecompl}

\usepackage{longtable}
\let\tablenum\relax
\usepackage{siunitx}

\usepackage{graphicx}
\usepackage{subfigure}
\usepackage{natbib}
\usepackage{amsmath}
\usepackage{lastpage}

\def\CIVdbl{{\rm C~}\kern 0.1em{\sc iv}~$\lambda\lambda 1548, 1550$}
\def\MgIIdbl{{\rm Mg~}\kern 0.1em{\sc ii}~$\lambda\lambda 2796, 2803$}
\def\NVdbl{{\rm N}\kern 0.1em{\sc v}~$\lambda\lambda 1238, 1242$}  
\def\OVIdbl{{\rm O}\kern 0.1em{\sc vi}~$\lambda\lambda 1031, 1037$}
\def\SiIVdbl{{\rm Si~}\kern 0.1em{\sc iv}~$\lambda\lambda1394, 1403$}
\def\AlIIIdbl{{\rm Al~}\kern 0.1em{\sc iii}~$\lambda\lambda1855,1863$}
\def\FeIIdbl{{\rm Fe~}\kern 0.1em{\sc ii}~$\lambda\lambda 2383, 2600$}

\def\AlII{\hbox{{\rm Al~}\kern 0.1em{\sc ii}}}
\def\AlI{\hbox{{\rm Al~}\kern 0.1em{\sc i}}}
\def\AlIII{\hbox{{\rm Al~}\kern 0.1em{\sc iii}}}
\def\CaI{\hbox{{\rm Ca}\kern 0.1em{\sc i}}}
\def\CaII{\hbox{{\rm Ca}\kern 0.1em{\sc ii}}}
\def\CrII{\hbox{{\rm Cr}\kern 0.1em{\sc ii}}}
\def\C{\hbox{{\rm C~}}}
\def\CI{\hbox{{\rm C~}\kern 0.1em{\sc i}}}
\def\CII{\hbox{{\rm C~}\kern 0.1em{\sc ii}}}
\def\CIII{\hbox{{\rm C~}\kern 0.1em{\sc iii}}}
\def\CIV{\hbox{{\rm C~}\kern 0.1em{\sc iv}}}
\def\CV{\hbox{{\rm C}\kern 0.1em{\sc v}}}
\def\H{\hbox{{\rm H}}}
\def\HI{\hbox{{\rm H~}\kern 0.1em{\sc i}}}
\def\HII{\hbox{{\rm H~}\kern 0.1em{\sc ii}}}
\def\Lya{\hbox{{\rm Ly}\kern 0.1em$\alpha$}}
\def\Lyb{\hbox{{\rm Ly}\kern 0.1em$\beta$}}
\def\Lyg{\hbox{{\rm Ly}\kern 0.1em$\gamma$}}
\def\Lyfive{\hbox{{\rm Ly}\kern 0.1em$5$}}
\def\Lysix{\hbox{{\rm Ly}\kern 0.1em$6$}}
\def\Lyseven{\hbox{{\rm Ly}\kern 0.1em$7$}}
\def\Lyeight{\hbox{{\rm Ly}\kern 0.1em$8$}}
\def\Lynine{\hbox{{\rm Ly}\kern 0.1em$9$}}
\def\Lyten{\hbox{{\rm Ly}\kern 0.1em$10$}}
\def\HeI{\hbox{{\rm He}\kern 0.1em{\sc i}}}
\def\HeII{\hbox{{\rm He}\kern 0.1em{\sc ii}}}
\def\FeI{\hbox{{\rm Fe~}\kern 0.1em{\sc i}}}
\def\FeII{\hbox{{\rm Fe~}\kern 0.1em{\sc ii}}}
\def\FeIII{\hbox{{\rm Fe~}\kern 0.1em{\sc iii}}}
\def\MnII{\hbox{{\rm Mn}\kern 0.1em{\sc ii}}}
\def\MgI{\hbox{{\rm Mg~}\kern 0.1em{\sc i}}}
\def\MgII{\hbox{{\rm Mg~}\kern 0.1em{\sc ii}}}
\def\MgIII{\hbox{{\rm Mg~}\kern 0.1em{\sc iii}}}
\def\MgIV{\hbox{{\rm Mg~}\kern 0.1em{\sc iv}}}
\def\NaI{\hbox{{\rm Na}\kern 0.1em{\sc i}}}
\def\NV{\hbox{{\rm N}\kern 0.1em{\sc v}}}
\def\NII{\hbox{{\rm N}\kern 0.1em{\sc ii}}}
\def\NIII{\hbox{{\rm N}\kern 0.1em{\sc iii}}}
\def\O{\hbox{{\rm O}}}

\def\OVI{\hbox{{\rm O}\kern 0.1em{\sc vi}}}
\def\OIV{\hbox{{\rm O}\kern 0.1em{\sc iv}}}
\def\OI{\hbox{{\rm O}\kern 0.1em{\sc i}}}
\def\OII{\hbox{{\rm O}\kern 0.1em{\sc ii}}}
\def\OIII{\hbox{{\rm O}\kern 0.1em{\sc iii}}}
\def\PV{\hbox{{\rm P}\kern 0.1em{\sc v}}}
\def\SiII{\hbox{{\rm Si~}\kern 0.1em{\sc ii}}}
\def\SiIII{\hbox{{\rm Si~}\kern 0.1em{\sc iii}}}
\def\SiIV{\hbox{{\rm Si~}\kern 0.1em{\sc iv}}}
\def\SII{\hbox{{\rm S}\kern 0.1em{\sc ii}}}
\def\SIII{\hbox{{\rm S}\kern 0.1em{\sc iii}}}
\def\SIV{\hbox{{\rm S}\kern 0.1em{\sc iv}}}
\def\SVI{\hbox{{\rm S}\kern 0.1em{\sc vi}}}
\def\TiII{\hbox{{\rm Ti}\kern 0.1em{\sc ii}}}
\def\ZnII{\hbox{{\rm Zn}\kern 0.1em{\sc ii}}}
\def\kms{\hbox{km~s$^{-1}$}}

\newcommand{\s}{$\sigma$}

\def\ang{$\text \AA$}

\def\lsim{\mathrel{\rlap{\lower4pt\hbox{\hskip1pt$\sim$}}
    \raise1pt\hbox{$<$}}}                
\def\gsim{\mathrel{\rlap{\lower4pt\hbox{\hskip1pt$\sim$}}
    \raise1pt\hbox{$>$}}}

\def\citeapos#1{\citeauthor{#1}'s (\citeyear{#1})}

\begin{document}

\title{Survey of Extremely-High-Velocity Outflows in Sloan Digital Sky Survey Quasars}

\correspondingauthor{Paola Rodr\'iguez Hidalgo}
\email{paola@uw.edu}

\author{Paola Rodriguez Hidalgo}
\affiliation{Physical Sciences Division, School of STEM, University of Washington Bothell, WA, 98011, USA}
\affiliation{Department of Physics and Astronomy, Humboldt State University, Arcata, CA 95521, USA}
\affiliation{Department of Physics and Astronomy, York University, Toronto, ON M3J 1P3, Canada}
\affiliation{Department of Astronomy \&\ Astrophysics, University of Toronto, Toronto, ON M5S 3H8, Canada}

\author{Abdul Moiz Khatri}
\affiliation{Department of Astronomy \&\ Astrophysics, University of Toronto, Toronto, ON M5S 3H8, Canada}

\author{Patrick B. Hall}
\affiliation{Department of Physics and Astronomy, York University, Toronto, ON M3J 1P3, Canada}

\author{Sean Haas}
\affiliation{Department of Physics and Astronomy, Humboldt State University, Arcata, CA 95521, USA}

\author{Carla Quintero}
\affiliation{Department of Physics and Astronomy, Humboldt State University, Arcata, CA 95521, USA}

\author{Viraja Khatu}
\affiliation{Department of Astronomy \&\ Astrophysics, University of Toronto, Toronto, ON M5S 3H8, Canada}
\affiliation{Department of Astronomy \&\ Astrophysics, Western University, London, ON N6A 3K7, Canada}

\author{Griffin Kowash}
\affiliation{Department of Physics and Astronomy, Humboldt State University, Arcata, CA 95521, USA}

\author{Norm Murray}
\affiliation{Canadian Institute for Theoretical Astrophysics, University of Toronto, Toronto, ON M5S 3H8, Canada}

\begin{abstract} 
We present a survey of extremely high-velocity outflows (EHVOs) in quasars, defined by speeds between 0.1$c$ and 0.2$c$. This region of the parameter space has not been included in previous surveys, but it might present the biggest challenge for theoretical models and it might be a large contributor to feedback due to the outflows' potentially large kinetic power. 
Using the Sloan Digital Sky Survey, we find 40 quasar spectra with broad EHVO {\CIV} absorption, 10 times more than the number of previously known cases. We characterize the EHVO absorption and find that in 26 cases, {\CIV} is accompanied by {\NV} and/or {\OVI} absorption. 
We find that EHVO quasars lack {\HeII} emission and have overall larger bolometric luminosities and black hole masses than those of their parent sample and BALQSOs, while we do not find significant differences in their Eddington ratios. We also report a trend toward larger black hole masses as the velocity of the outflowing gas increases in the BALQSOs in our sample. The overall larger $L_{\rm bol}$ and lack of {\HeII} emission of EHVO quasars suggest that radiation is likely driving these outflows. We find a potential evolutionary effect as EHVO quasars seem to be more predominant at large redshifts. We estimate that the kinetic power of these outflows may be similar to or even larger than that of the outflows from BALQSOs as the velocity factor increases this parameter by 1--2.5 orders of magnitude. Further study of EHVO quasars will help improve our understanding of quasar physics.
\end{abstract}

 
\section{Introduction}

Quasars, the most luminous of the active galactic nuclei (AGNs), are found at the center of most massive galaxies. A quasar's luminosity is generated by an accretion disk surrounding a supermassive black hole. Collisions in the gas disk cause matter to slowly spiral into the black hole, and the gravitational potential energy released by that gas is emitted from the disk as electromagnetic radiation. 
Quasars' large luminosities allow us to study them at large redshifts, providing information about galactic evolution within our universe. 

Outflows are fundamental constituents of AGNs and they provide first-hand information about the physical and chemical properties of the AGN environment. Quasar outflows can be caused by material outflowing from the accretion disk, and they are detected in a substantial fraction of AGNs through absorption-line signatures (e.g., broad, blue-shifted resonance lines in the UV and X-ray bands) as the gas intercepts some of the light from the central continuum source and broad emission-line region (e.g. \citealt{Crenshaw99}; \citealt{Reichard03}; \citealt{Hamann04}; \citealt{Trump06}; \citealt{Dunn08}; \citealt{Ganguly08}; \citealt{Nestor08} and references therein). Outflows could be ubiquitous, though, if the absorbing gas subtends a small solid angle around the background source. Outflows have been invoked as a potentially regulating mechanism that would provide the necessary energy and momentum ``feedback'' \citep[e.g.,][]{Silk98, DiMatteo05, Springel05, Hopkins06} required to explain the correlation between the black hole masses ($M_{\rm BH}$) and the masses of the stellar spheroids ($M_{\rm bulge}$) of their host galaxies  \citep[e.g.,][]{Gebhardt00, Merritt01, Tremaine02}.

In particular, gas outflowing at extremely high speeds might be the most disruptive to the host galaxy environment, due to their large kinetic power.
Outflows with speeds $v\sim$0.2$c$ carry approximately 1-2.5 orders-of-magnitude larger kinetic power than gas outflowing at what is defined as "high" velocities ($v \sim$5,000--10,000 {\kms}), if their gas is located at similar distances and has similar physical properties, because kinetic power is proportional to $v^3$. 
Extremely high-velocity outflows (EHVOs) might also pose the biggest challenges to theoretical models that try to explain how these outflows are launched and driven (\citealt{Hamann02}; \citealt{Sabra03}). Radiation pressure models (\citealt{Arav94}; \citealt{Murray95}; \citealt{Proga00}; \citealt{Ostriker10}; and the excellent review in \citealt{Crenshaw03}) invoke the powerful central source to accelerate line-driven winds and have proven successful in explaining different aspects of these winds, such as the relation between the AGN luminosity and the terminal velocity of the outflow (\citealt{Laor02}) and ``line locking'' (\citealt{Turnshek88}; \citealt{Srianand02}; \citealt{Hamann10}). However, simulations and theoretical models have not yet shown the presence of detached profiles with central velocities as large as 0.2$c$. Alternatively, some models have also invoked magnetic forces to launch, drive, and constrain the flow (\citealt{deKool95}; \citealt{Proga04}; \citealt{Everett05}), and higher terminal velocities are expected in magnetic driving, due to stronger centrifugal forces (\citealt{Proga07}).

Outflows with speeds larger than 0.1$c$ have been detected as UV/optical absorption lines in individual quasars (e.g., \citealt{Januzzi96}; \citealt{Hamann97a}; \citealt{RodriguezHidalgo11}; \citealt{Rogerson16}). Additionally, ultrafast outflows (UFOs) have been observed as Fe K-shell absorption in the X-ray spectra of mostly nearby AGNs (predominantly Seyferts) at speeds up to 0.4$c$ (e.g., \citealt{Chartas02}; \citealt{Reeves03}; \citealt{Tombesi10}; and references therein). UFO absorption is typically narrow and weak (EW $\leq $ 150 eV), and large systematic studies are prohibitive and restricted to nearby AGNs. 

EHVOs in UV/optical spectra have been barely studied systematically prior to this work.  For simplicity, large UV/optical surveys of quasar outflows have focused on searching for broad {\CIV} absorption 
that would indicate gas outflowing at speeds less than 0.1$c$. Among different ionic transitions, {\CIV} $\lambda\lambda$1548.1950,1550.7700 
is (1) commonly present in quasar outflows and (2) easily observed due to the fact that it is redshifted into the optical range for quasars in the epoch of peak quasar activity (for luminous Type 1 quasars, the comoving space density peaked at redshifts 2$< z_{\rm em} <$3; \citealt{Schmidt95}; \citealt{Ross13} and references therein). The arbitrary velocity limit of 0.1$c$ is defined to avoid complications due to misidentification with {\SiIV} or other ionic transitions  blue-shifted of the {\SiIV} emission line. However, doubling the speed results in almost one order-of-magnitude increase in the kinetic power (assuming all other physical parameters remain similar), 
so failing to account for EHVOs may lead to underestimates of the impact of quasar feedback on galaxies.

With the goal of creating the first database of EHVOs, we searched for broad (widths larger than 1000 {\kms}) and non-shallow (depths larger than 10\% of the normalized flux) {\CIV} absorption that appears blue-shifted at speeds of 0.1$c$--0.2$c$ in quasar spectra. In this paper, we present the results of this search carried out over quasars in the ninth release of the Sloan Digital Sky Survey (SDSS) quasar catalog (DR9Q; \citealp{Paris12}), which is derived from the Baryon Oscillation Spectroscopic Survey (BOSS) of SDSS-III. In Section \ref{sec:data}, we describe the parent sample. In Section \ref{sec:search}, we explain how we normalized the spectra and searched for EHVOs seen in absorption in quasar spectra. In Section \ref{sec:results}, we present the results of this search, the properties of the absorption and of the EHVO quasars themselves relative to the parent and BALQSO samples, and the identification of outflows in other ionic transitions at extremely high speeds. We discuss, as well, some preliminary analysis of {\HeII} emission and radio loudness in EHVO quasars.  Finally, in Section \ref{sec:discussion} we discuss how EHVO outflows compare to other classes of quasar outflows and the implications that these results have for outflow driving mechanisms and for feedback on the host galaxies of these outflows. 

\section{Data -- Original and parent samples}
\label{sec:data}

The ionic transition of {\CIV} in gas outflowing at speeds from 0.1$c$ to 0.2$c$ is blueshifted into the region between the Ly{$\alpha$} and {\SiIV+\OIV]} emission lines of the quasar spectrum. 
This region is observed in the optical part of the spectrum in quasars at the peak of the quasar activity epoch (for luminous Type 1 quasars, the comoving space density peaked at redshifts 2$< z_{\rm em} <$3; \citealt{Schmidt95}; \citealt{Ross13} and references therein).
The SDSS (\citealt{York00})
has released several quasar catalogs with publicly accessible optical quasar spectra,
so it has become the optimal survey in which to search for {\CIV} absorption (e.g., \citealt{Paris12}).\\

The original sample for our work was the ninth release of the SDSS quasar catalog \citep[DR9Q;][]{Paris12} which was derived from the
BOSS (\citealt{Dawson13})
of SDSS-III \citep{Eisenstein11}.
DR9Q contains a total of 87,822 quasar spectra and has wavelength coverage from 3600 to 10500 \ang. DR9Q includes a variety of useful measurements, such as emission redshifts and the signal-to-noise ratio (S/N) for each quasar spectrum (\citealt{Paris12}).  \\

\par
To construct a parent sample from which to search for EHVO quasars, we performed two cutoffs:
\begin{itemize}
\item[1.] to search for {\CIV} outflowing at speeds up to 0.2$c$, we only included quasars with emission redshift $z_{\rm em} \geq {1.9}$. This places the Ly{$\alpha$} emission line within the BOSS wavelength coverage in all cases. For this $z_{\rm em}$ selection, we used DR9Q measurements of $z_{\rm PCA}$, 
\item[2.] to avoid spurious and ambiguous detections, we only included quasar spectra with S/N $>$ 10. 
(For comparison, \citet{Gibson09b} used a ``$SNR\_1700 \geq $ 9' when defining what they called a  ``high-SNR subsample'' of BALQSOs.)
For this selection, we used the DR9Q measurements of ``SNR\_1700'', which included the closest rest-frame wavelength window (1650 -- 1750 \AA) to our wavelengths of interest.  
\end{itemize}

After imposing these two cutoffs, our parent sample included 6760 quasar spectra. These two cutoffs resulted in a sample of quasars where EHVOs with the characteristics explained in Section \ref{search_abs} could be detected unequivocally. 
Because there is only a difference of $\sim$400 {\AA} between the region of interest and the centroid of this measurement provided by SDSS, the S/N value in the region of interest tends to be very similar to the one provided by SDSS, confirmed by our own measurements of the S/N. Typically, the normalized error spectrum is quite flat (see the quasar spectra in Section \ref{sec:search}). Only for quasars with 1.9 $ < z_{\rm em} < $ 2 might the S/N decrease more rapidly in the wavelength region of interest. At those redshifts, this spectral region is close to the blue edge of the SDSS spectra where the sensitivity is poorer. However, less than 1\% of our sample (65/6760) shows a combination of low redshift and low $SNR\_1700$. Those cases were carefully inspected and rejected if necessary (see Section \ref{sec:search}).

\section{Searching for {\CIV} Extremely High-velocity Outflows}
\label{sec:search}

To search systematically for EHVO {\CIV} absorption in our parent sample, we (1) normalized all spectra, (2) searched for any absorption appearing below the normalized continuum and stronger than 10\% of the continuum value, and (3) rejected any other possible identifications for the ionic transition (such as {\SiIV}, , or {\CII}). Each of these steps is explained further below. 

\subsection{Normalization of the quasar spectra}
\label{norm}

\begin{figure}
\centering
\includegraphics[width=8.25cm]{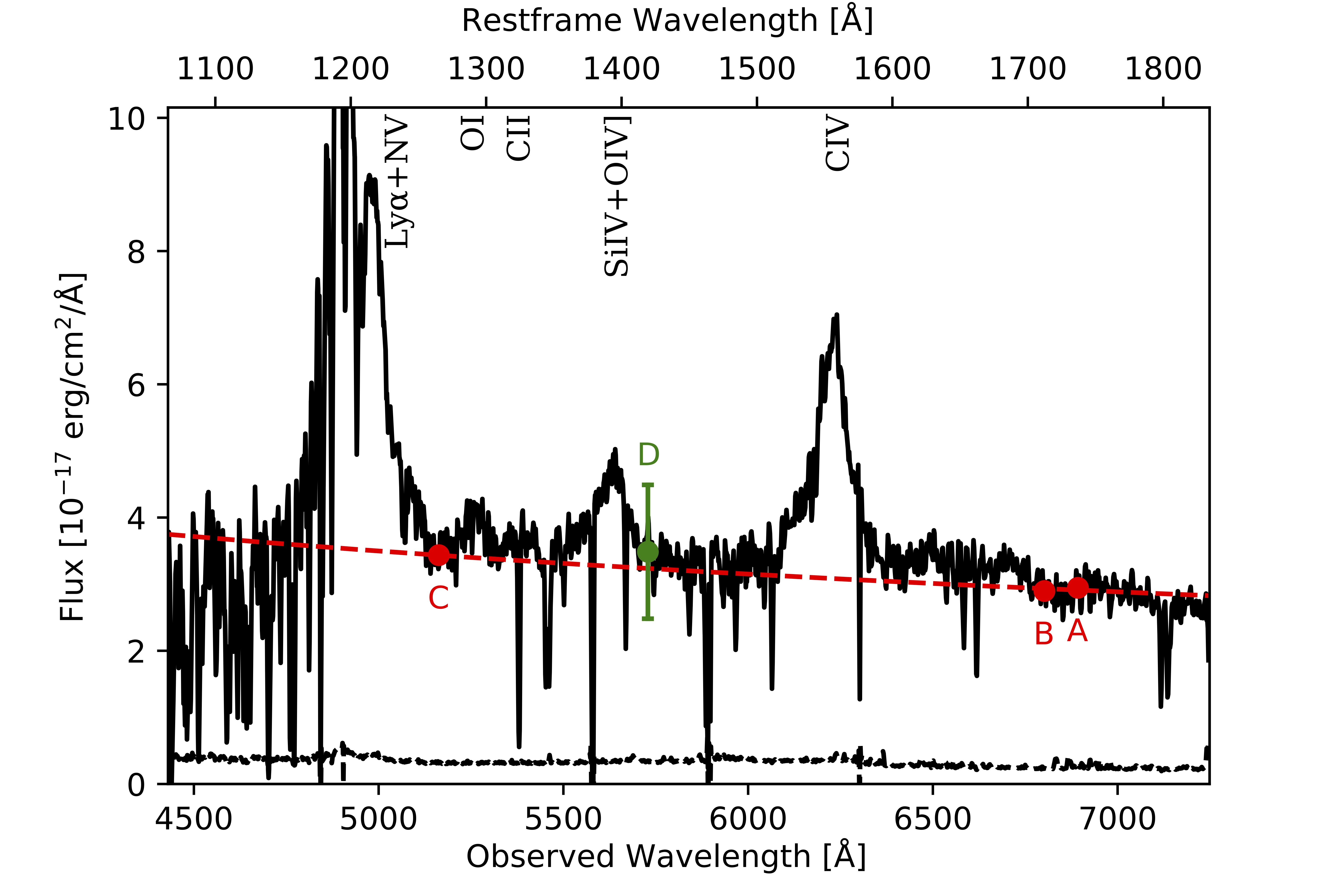}
\caption{Example of the normalization explained in Section \ref{norm}.  We fitted a power law (red dashed line) to the original spectrum of each quasar (black solid line). The power law is constructed by using the points A (median flux of points with rest-frame wavelengths between 1701 - 1725 \textup{\AA}), B (1677 - 1701 \textup{\AA}), and C (1280 - 1284 \textup{\AA}). Point D (1415 - 1430 \textup{\AA}) was used to determine whether the power-law fit deviated more than three times the median error from the spectrum at that location and was used instead of C when that was the case. Selected emission lines present in the spectrum are labeled at the top, and both the observed and quasar rest-frame wavelength are shown (bottom and top, respectively).} 
\label{SDSSnorm}
\end{figure}

To search systematically for {\CIV} absorption that appears below the quasar continuum, we started by normalizing the quasar spectra.
We have taken a conservative normalization approach: we selected the lowest possible fit through the continuum; thus, our absorption measurements might be lower limits in those cases where we underfit the continuum if the location of the ``true'' continuum is uncertain. Figure \ref{SDSSnorm} shows an example of our systematic normalization. The majority of quasars show greater flux at shorter wavelengths
in the UV/optical region, and composite quasar spectra are well approximated by a power law in this region, even in the far UV (e.g, \citealp{VandenBerk01}, \citealt{Telfer02}, \citealt{Tilton16}). We used a simple power law (similar to what is shown in Figure 6 of \citealp{VandenBerk01}) anchored to three points in the quasar spectra (see Figure \ref{SDSSnorm}). 
These points are defined as the central value of the wavelength and the median value of the flux in four distinct wavelength regions. Regions A (rest-frame 1701-1725 \ang) and B (rest-frame 1677-1701 \ang) are located in spectral regions where we determined emission and absorption are typically not present after visual inspection and comparison to emission-line tables. Regions C (rest-frame 1280-1284 \ang) and D (rest-frame 1415-1430 \ang) were used to define the slope of the power law. As both C and D may sometimes be affected by emission and/or absorption, we adopted the following algorithm: (1) we used region C, together with A and B, to define the power law;
(2) we compared this power-law fit at the midpoint of region D with the median value of the flux in region D; (3) if the difference between the two was not zero within 3$\sigma$ (three times the median error in region D), we used region D, together with A and B, to define the power law instead. 
We ignored bad pixels with zero inverse variance.

Besides the possibility of absorption being present, the spectral region between the Ly{$\alpha$} and {\SiIV+\OIV]} emission lines may be complicated by additional emission lines that are only sometimes present and often weaker than {\SiIV+\OIV]} (such as {\OI} and {\CII}; see Figure \ref{SDSSnorm}).\footnote{See \citealt{RodriguezHidalgo11} for an example of a deblending analysis of the emission lines in this region.} In this paper, we did not attempt to fit these emission lines for each individual spectrum, 
but all normalizations were visually inspected.

For those cases where the systematic fit was judged to be unsuccessful (e.g., too low), we attempted individual normalizations, substituting or including additional anchor points throughout the whole spectrum. 
When the continuum between the Ly{$\alpha$} and {\SiIV+\OIV]} emission lines was complex, we kept the spectrum in our sample only if the individually normalized fit at 1415 - 1725 \AA\ was successful 
at defining a plausible continuum when extrapolated into this complex region; five spectra in which this was not the case were rejected from our sample. Additionally, we removed 12 spectra that lacked flux information for more than half the wavelength region of interest. In total, we rejected 17 cases, reducing the parent sample to 6743 quasar spectra.

\subsection{Absorption Detection}  
\label{search_abs}

The next step consisted in identifying any broad and non-shallow absorption features in the spectral region corresponding to {\CIV} absorption outflowing at velocities between 0.1$c$ and 0.2$c$. We restricted our search to cases where the continuum was absorbed: because we do not fit a pseudo-continuum to the continuum+emission lines (such as in \citealt{RodriguezHidalgo13}), we only detect absorption below the continuum level. 
To search for absorption, 
we smoothed the spectra with a three-pixel boxcar.
Measurements of absorption parameters described in Section \ref{sec:results}, though,  were carried out over the unsmoothed spectra.

{{\it Why broad absorption?}}
Quasar spectra can include both absorption in the vicinity of the quasar (typically known as "intrinsic") and absorption that lies along the line of sight between the quasar and us, but is located outside the quasar environment ("intervening" absorption). Intrinsic absorption is not distinguishable from intervening narrow absorption at the SDSS resolution with only one observation (high-resolution observations and variability studies can be used to identify their nature, however; \citealt{Narayanan04}; \citealt{Misawa07a}). Intervening absorbers in rich and massive clusters of galaxies (e.g., \citealt{Girardi98}; \citealt{Venemans07}) show that the velocity dispersion of intervening absorbers can reach up to $\sim$1000--1200 {\kms}. Therefore, the broader the absorption, the more likely it is intrinsically related to the supermassive black hole phenomenon. To avoid intervening absorption, we searched for broad absorption with widths larger than 1000 {\kms}. 

{{\it Why not broad but shallow absorption?}} If quasar spectra include broad and shallow absorption, our process for continuum normalization would fit through the absorption as if it were the continuum and the absorption would be undetectable. In order to detect shallow absorption with long widths, alternative methods for continuum normalization must be used (i.e., fitting quasar spectra templates). Within the method we used, we are not able to detect this type of absorption.

The Balnicity index (BI; \citealt{Weymann91}) 
is one of the standard ways to search systematically for and characterize absorption. The BI represents an equivalent-width measurement: it is larger when the absorption has larger width and/or depth, and it is calculated by the following integral: 
\begin{equation}
\centerline { BI $=  -\int_a^b \ [1 - \frac{f(v)}{0.9}]C \, \mathrm{d}v. $}\\
\end{equation} 
where $f(v)$ is the normalized flux (see \citealt{Weymann91}). Dividing $f(v)$ by 0.9 avoids detections of shallow absorption; the square bracket is only positive if the normalized flux values are lower than 0.9, thus absorption only contributes toward the BI if it reaches depths below 90\% of the continuum flux.
The $C$ variable only holds two values: zero and one. Initially, it is set to zero and only becomes one when the term in the square bracket is continuously positive over a velocity interval of our choice. To find only broad absorption, we set this velocity width to be 1000 {\kms} (see above). The values of $a$ and $b$ are the velocity integration limits. Typically, those limits are set to search only at velocities lower than 25,000 {\kms} to avoid contaminating the BI measurement derived from the {\CIV} absorption with other absorption that would typically accompany the {\CIV} absorption, such as {\SiIV}.
The BI definition has its limitations, and different values of $C$, $a$ and $b$ have also been proposed and used in different studies (see for example, \citealt{Hall02}; \citealt{Trump06}; \citealt{Gibson09b}). In our study, we set $a$ and $b$ to be 30,000 and 60,000 {\kms}, respectively, to flag any potential EHVO absorption. (We later determine whether the absorption is due to {\CIV} or other ions by visually inspecting each spectrum; see Section \ref{sec:reject}.) BI values can be slightly contaminated with intervening and unrelated absorption, but given the much larger width of {\CIV} intrinsic absorption, this represents a small portion of the total BI or equivalent-width measurement.

\subsection{Selection of {\CIV} EHVO and Rejection of Other Possible Identifications}
\label{sec:reject}

We calculated the BI as described above for all normalized spectra. For all spectra in which the BI was larger than zero, we carried out a careful visual inspection to select a secure EHVO sample. Potential reasons for rejecting spectra from the secure sample are explained below. 

First, the broad absorption might not be due to {\CIV}. While broad absorption appearing between the {\SiIV+\OIV]} and {\CIV} emission lines is typically attributed to the ionic transition of {\CIV}, absorption appearing between the Ly{$\alpha$} and {\SiIV+\OIV]} emission lines can be caused instead by (1) {\SiIV} absorption, outflowing at lower velocity, which is the most frequent occurrence of an alternative identification, or (2) {\OI}, {\SiII}, and {\CII} absorption, also outflowing at lower velocities.
In order to discriminate between other possible identifications and include in our sample only secure cases of extremely high-velocity outflowing {\CIV}, we visually inspected every spectrum where absorption was detected and follow the methodology described below. Given the large widths of the absorption features, notice that it is not possible to use the difference in doublet separation to identify ionic transitions. 

\begin{figure}
\centering
\includegraphics[width=8.25cm]{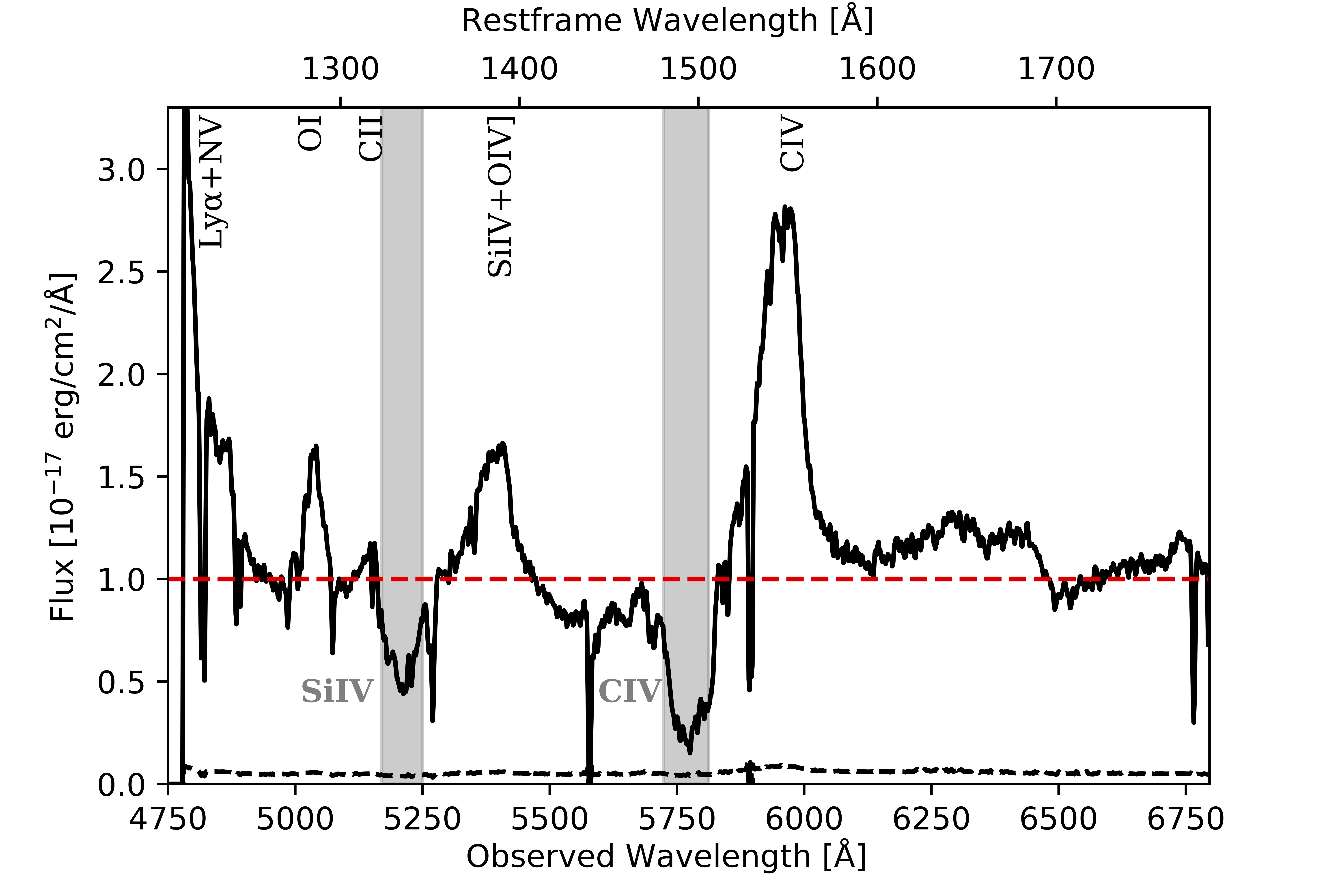}
\caption{Example of a rejected case as discussed in Section \ref{sec:reject}.
We include a dashed red line to indicate the normalized continuum level and help guide the eye. While there is absorption present with depth more than 10\% below the continuum in the wavelength region of interest, it is {\SiIV} outflowing at the same speeds as the corresponding {\CIV} (both shaded gray and labeled). Our code would flag the absorption, and it would be rejected by visual inspection. Similar to Figure \ref{SDSSnorm}, some emission lines present in the spectrum are labeled at the top.} 
\label{fig:rej}
\end{figure}

In quasar spectra, absorption due to {\SiIV} can be rejected easily in most cases because {\SiIV} absorption always has corresponding {\CIV} absorption outflowing at similar speeds; there are no confirmed cases in the literature of quasar outflows where {\SiIV} absorption is present without corresponding {\CIV}. This agrees with the solar Si/C abundance (\citealt{Grevesse98}; \citealt{Asplund09}; \citealt{Lodders10}), 
and the similar ionization potentials of {\SiIV} and {\CIV}.
We used a similar approach to determine any other possible identifications, given that {\SiIV} and {\CIV} are the predominant ions in these outflows. 

Figure \ref{fig:rej} shows an example where flagged absorption was rejected by visual inspection because it is {\SiIV} and not {\CIV}. All cases with flagged absorption were checked for potential alternative identifications by locating the wavelengths in the spectrum where the corresponding {\CIV} absorption outflowing at similar speeds would be. Whenever the identification of absorption in the region of interest was ambiguous and could be attributed to any other ion accompanied by {\CIV}, we discarded it and did not include it in our sample. This, again, will result in our survey being a conservative lower limit of the number of EHVO {\CIV} in quasar spectra. 

Another possible reason for rejection was the blending of narrow absorption that can result in apparent broad absorption. During visual inspection, we rejected any absorption profiles shapes that resembled blended narrow absorption. Intervening damped-Ly$\alpha$ absorption would thus be excluded from our sample. 

\begin{figure}
\centering
\includegraphics[width=7.7cm]{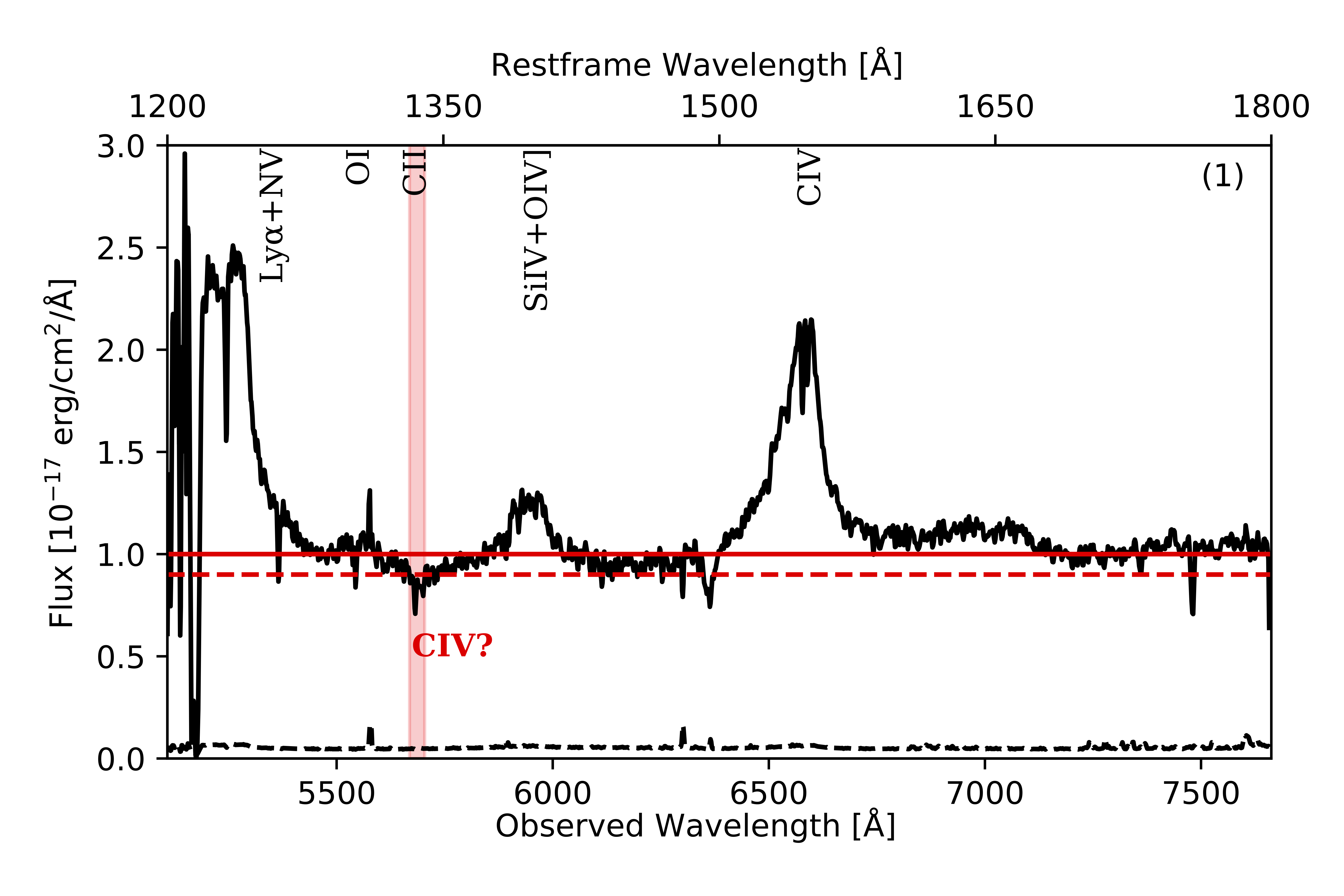}
\includegraphics[width=7.7cm]{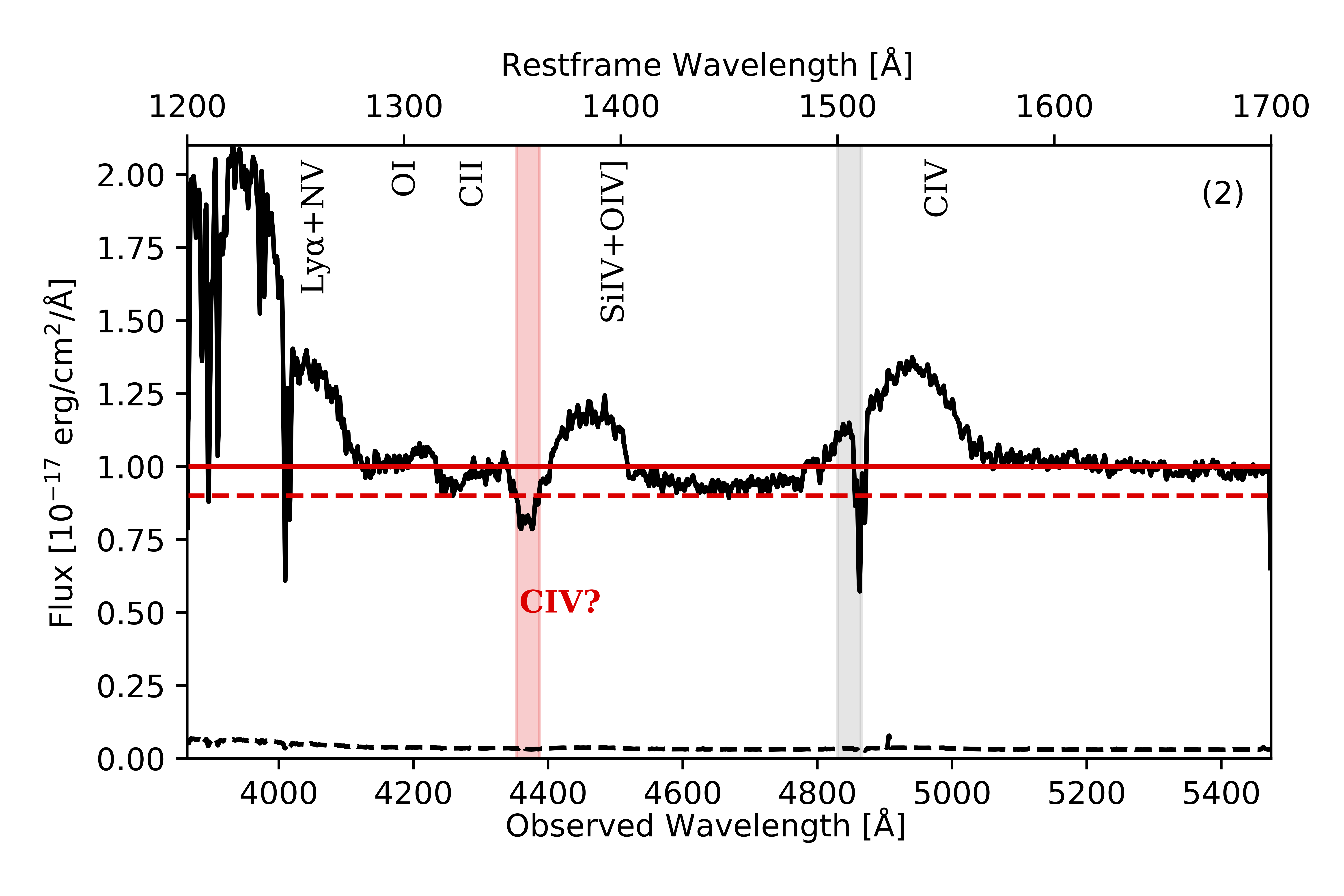}
\includegraphics[width=7.7cm]{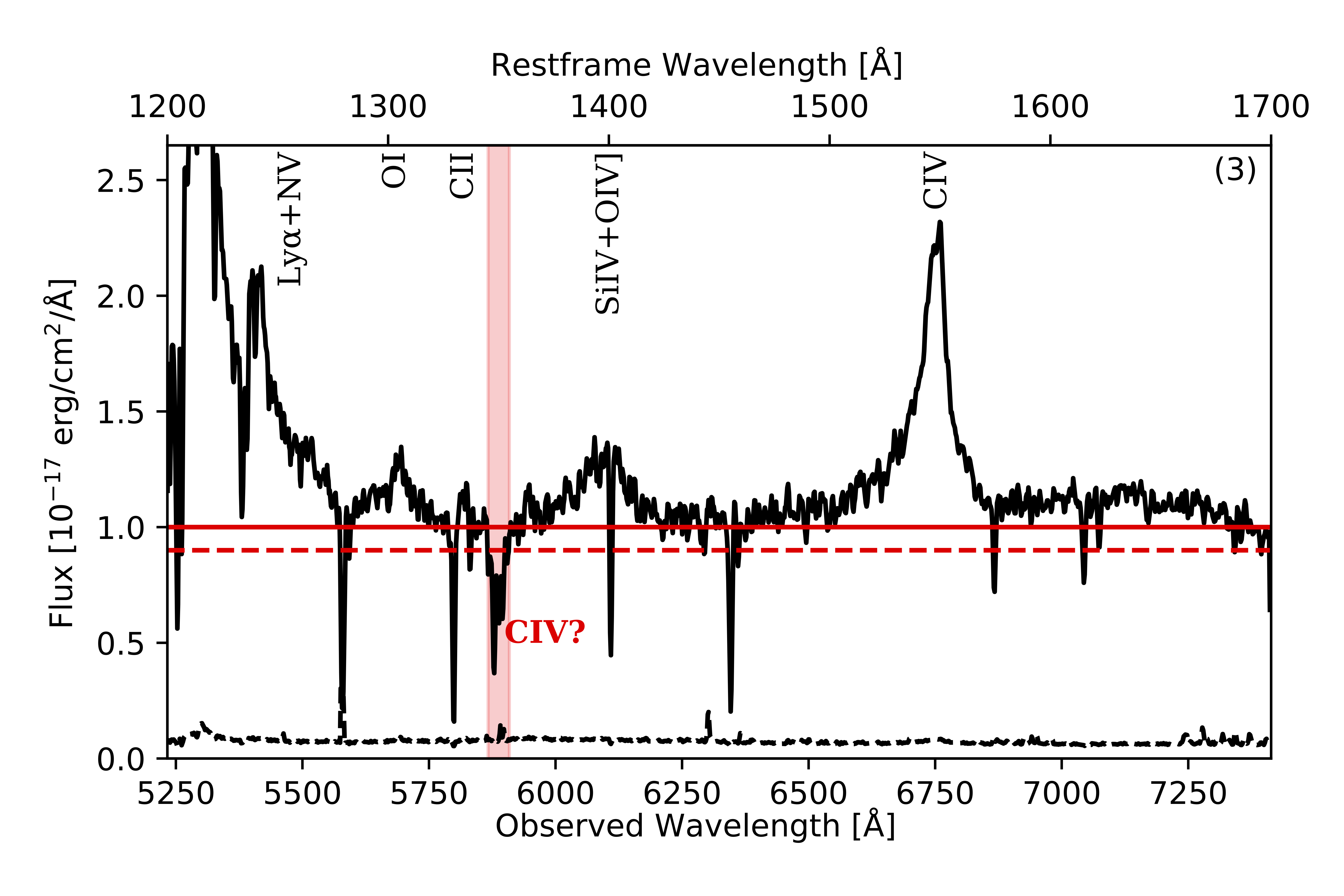}
\caption{Examples of candidates of quasars with potential EHVO {\CIV} absorption that were not included in our sample. The candidate {\CIV} absorption is indicated by a red shaded region whose limits are determined by the wavelengths where the absorption crosses the 90\% level of the normalized flux. In the top figure (1), we show an example where a different continuum placement would not have resulted in flagged absorption due to the weak nature of the absorption and the difficulty of placing the continuum. The middle panel (2) shows an example where the absorption could be due to {\SiIV} instead, because there is corresponding {\CIV} for a large portion of the trough's extent in velocity although the profiles do not mimic each other. The bottom panel (3) shows an example of a potential blend of narrow lines. Similar to Figure \ref{SDSSnorm}, some emission lines present in the spectrum are labeled at the top of each figure.}
\label{fig:CandidateExample}
\end{figure}

While many of these rejections were unequivocal, we found approximately 60 cases that potentially could be EHVO but that did not meet our standards for the current sample included in this paper (see Section \ref{sec:results}). Figure \ref{fig:CandidateExample} shows examples of each type of these {``{\it candidates}''}: these cases were ambiguous due to (1 - top panel in Figure \ref{fig:CandidateExample}) a combination of a difficult continuum placement and borderline depth or width of the absorption, which in combination results in cases where a different normalization would not result in flagged absorption. In some cases (2 - middle panel in Figure \ref{fig:CandidateExample}), the identification could be {\SiIV} due to the presence of {\CIV} in a large range of  corresponding velocities, although the profiles do not resemble each other so we suspect the absorption is not {\SiIV}. Finally (3 - bottom panel in Figure \ref{fig:CandidateExample}), some cases appear to be the result of either the blending of real broad absorption with narrow absorption or severe blending of only narrow absorption lines. 
We have not included any of these cases in our analysis below, but they will be available in our database. Our plan is to monitor these cases for potential future inclusion in our sample.

\begin{deluxetable*}{ccccc}
\tablecolumns{5} 
\tablecaption{Information about the EHVO Quasar Sample}
\tablewidth{40pt}
\tablehead{
\colhead{QSO} & \colhead{\hspace{1.15cm}Plate-MJD-fiber} & \colhead{\hspace{1.15cm}$z_{\rm em,DR9Q}$} & \colhead{\hspace{1.15cm}$z_{\rm em,HW10}$} & \colhead{\hspace{1.15cm}BALQSO?}
}
\startdata
J000154.90$-$004956.4 & {\hspace{1.15cm}}4216-55477-0166 & {\hspace{1.15cm}}2.8302$\pm$0.0005 & {\hspace{1.15cm}}... & {\hspace{1.15cm}}no \\ 
J001306.14+000431.8 & {\hspace{1.15cm}}4217-55478-0020 & {\hspace{1.15cm}}2.1643$\pm$0.0004 & {\hspace{1.15cm}}2.169$\pm$0.002 & {\hspace{1.15cm}}no \\  
J005922.65+000301.4 & {\hspace{1.15cm}}3735-55209-0532 & {\hspace{1.15cm}}4.1822$\pm$0.0013 & {\hspace{1.15cm}}4.177$\pm$0.008  &{\hspace{1.15cm}}no \\ 
J012700.69$-$004559.1 & {\hspace{1.15cm}}4229-55501-0337 & {\hspace{1.15cm}}4.0826$\pm$0.0027 & {\hspace{1.15cm}}4.097$\pm$0.002 & {\hspace{1.15cm}}no \\ 
J014548.55$-$000812.5 & {\hspace{1.15cm}}4231-55444-0035 & {\hspace{1.15cm}}2.7903 & {\hspace{1.15cm}}2.805$\pm$0.002 & {\hspace{1.15cm}}yes \\ 
J071843.50+391720.3 & {\hspace{1.15cm}}3655-55240-0388 & {\hspace{1.15cm}}2.6298$\pm$0.0006 & {\hspace{1.15cm}}... & {\hspace{1.15cm}}no \\
J073505.97+280327.0 & {\hspace{1.15cm}}4456-55537-0704 & {\hspace{1.15cm}}2.3243 & {\hspace{1.15cm}}... & {\hspace{1.15cm}}no \\
J074711.14+273903.3 & {\hspace{1.15cm}}4452-55536-0214 & {\hspace{1.15cm}}4.1143$\pm$0.0011 & {\hspace{1.15cm}}4.128$\pm$0.003 & {\hspace{1.15cm}}no \\ 
J075240.18+092523.0 & {\hspace{1.15cm}}4511-55602-0415 & {\hspace{1.15cm}}3.0223$\pm$0.0007 & {\hspace{1.15cm}}... & {\hspace{1.15cm}}no \\
J075852.68+133530.8 & {\hspace{1.15cm}}4506-55568-0824 & {\hspace{1.15cm}}3.3734$\pm$0.0014 & {\hspace{1.15cm}}3.386$\pm$0.007 & {\hspace{1.15cm}}no \\ 
J081337.14+155705.4 & {\hspace{1.15cm}}4498-55615-0502 & {\hspace{1.15cm}}2.4498$\pm$0.0007 & {\hspace{1.15cm}}... & {\hspace{1.15cm}}no \\
J083304.73+415331.3 & {\hspace{1.15cm}}3808-55513-0892 & {\hspace{1.15cm}}2.3583$\pm$0.0008 &  {\hspace{1.15cm}}2.329$\pm$0.007 & {\hspace{1.15cm}}no \\ 
J085825.71+005006.7 & {\hspace{1.15cm}}3815-55537-0910 & {\hspace{1.15cm}}2.8684$\pm$0.0007 & {\hspace{1.15cm}}2.863$\pm$0.003 & {\hspace{1.15cm}}no \\ 
J092125.97$-$023411.9 & {\hspace{1.15cm}}3766-55213-0046 & {\hspace{1.15cm}}2.8047$\pm$0.0005 & {\hspace{1.15cm}}... & {\hspace{1.15cm}}no \\ 
J094023.55+404703.2 & {\hspace{1.15cm}}4571-55629-0730 & {\hspace{1.15cm}}2.5839$\pm$0.0007 & {\hspace{1.15cm}}... & {\hspace{1.15cm}}no \\
J094258.03+005359.5 & {\hspace{1.15cm}}3826-55563-0860 & {\hspace{1.15cm}}2.4749 & {\hspace{1.15cm}}... & {\hspace{1.15cm}}no \\ 
J095005.90+362455.2 & {\hspace{1.15cm}}4573-55587-0698 & {\hspace{1.15cm}}2.2442$\pm$0.0004 & {\hspace{1.15cm}}2.237$\pm$0.002 & {\hspace{1.15cm}}no \\
J095254.10+021932.8 & {\hspace{1.15cm}}4743-55645-0118 & {\hspace{1.15cm}}2.1475 & {\hspace{1.15cm}}2.153$\pm$0.002 & {\hspace{1.15cm}}no \\ 
J095603.47+382517.2 & {\hspace{1.15cm}}4570-55623-0296 & {\hspace{1.15cm}}2.8962$\pm$0.0009 & {\hspace{1.15cm}}... & {\hspace{1.15cm}}yes \\  
J100400.20+372551.9 & {\hspace{1.15cm}}4567-55589-0454 & {\hspace{1.15cm}}2.5347 & {\hspace{1.15cm}}... & {\hspace{1.15cm}}yes \\ 
J103456.31+035859.4 & {\hspace{1.15cm}}4772-55654-0106 & {\hspace{1.15cm}}3.3867$\pm$0.0007 & {\hspace{1.15cm}}3.388$\pm$0.002 & {\hspace{1.15cm}}no \\ 
J110841.93+031735.1 & {\hspace{1.15cm}}4741-55704-0786 & {\hspace{1.15cm}}3.3478$\pm$0.0009 & {\hspace{1.15cm}}... & {\hspace{1.15cm}}yes \\  
J111154.35+372321.2 & {\hspace{1.15cm}}4622-55629-0678 & {\hspace{1.15cm}}2.0747$\pm$0.0003 & {\hspace{1.15cm}}2.074$\pm$0.002 & {\hspace{1.15cm}}no \\ 
J113000.22+344625.9 & {\hspace{1.15cm}}4619-55599-0854 & {\hspace{1.15cm}}3.6188$\pm$0.0008 & {\hspace{1.15cm}}3.615$\pm$0.004 & {\hspace{1.15cm}}no \\ 
J120609.69+004522.6 & {\hspace{1.15cm}}3844-55321-0920 & {\hspace{1.15cm}}2.6905$\pm$0.0006 & {\hspace{1.15cm}}2.691$\pm$0.003 & {\hspace{1.15cm}}no \\ 
J123056.28+345201.7 & {\hspace{1.15cm}}3968-55590-0464 & {\hspace{1.15cm}}3.4913$\pm$0.0010 & {\hspace{1.15cm}}3.493$\pm$0.003 & {\hspace{1.15cm}}no \\ 
J131307.93+065349.0 & {\hspace{1.15cm}}4840-55690-0576 & {\hspace{1.15cm}}2.9136 & {\hspace{1.15cm}}... & {\hspace{1.15cm}}no \\ 
J133150.25+393417.7 & {\hspace{1.15cm}}4708-55704-0418 & {\hspace{1.15cm}}1.9996$\pm$0.0005 & {\hspace{1.15cm}}1.991$\pm$0.003 & {\hspace{1.15cm}}yes \\ 
J133752.36$-$011924.7 & {\hspace{1.15cm}}4046-55605-0085 & {\hspace{1.15cm}}2.7999 & {\hspace{1.15cm}}2.835$\pm$0.005 & {\hspace{1.15cm}}yes \\  
J135203.03+333938.3 & {\hspace{1.15cm}}3861-55274-0582 & {\hspace{1.15cm}}3.0742$\pm$0.0014 & {\hspace{1.15cm}}3.079$\pm$0.004 & {\hspace{1.15cm}}no \\ 
J144356.21+062539.2 & {\hspace{1.15cm}}4858-55686-0392 & {\hspace{1.15cm}}2.6063$\pm$0.0006 & {\hspace{1.15cm}}... & {\hspace{1.15cm}}yes \\  
J150339.76+183423.9 & {\hspace{1.15cm}}3957-55664-0568 & {\hspace{1.15cm}}2.1848$\pm$0.0003 & {\hspace{1.15cm}}2.184$\pm$0.002 & {\hspace{1.15cm}}no \\ 
J151016.40+034200.0 & {\hspace{1.15cm}}4776-55652-0216 & {\hspace{1.15cm}}2.2468$\pm$0.0006 & {\hspace{1.15cm}}... & {\hspace{1.15cm}}no \\
J162445.03+271418.7 & {\hspace{1.15cm}}5006-55706-0846 & {\hspace{1.15cm}}4.4572$\pm$0.0033 & {\hspace{1.15cm}}4.478$\pm$0.003 & {\hspace{1.15cm}}no \\ 
J162747.14+192639.7 & {\hspace{1.15cm}}4060-55359-0940 & {\hspace{1.15cm}}2.4541$\pm$0.0006 & {\hspace{1.15cm}}... & {\hspace{1.15cm}}no \\
J164653.72+243942.2 & {\hspace{1.15cm}}4181-55685-0543 & {\hspace{1.15cm}}3.0329$\pm$0.0006 & {\hspace{1.15cm}}3.040$\pm$0.002 & {\hspace{1.15cm}}no \\ 
J165436.85+222733.8 & {\hspace{1.15cm}}4178-55653-0608 & {\hspace{1.15cm}}4.6910$\pm$0.0032 & {\hspace{1.15cm}}4.708$\pm$0.003 & {\hspace{1.15cm}}no \\ 
J171800.39+314203.9 & {\hspace{1.15cm}}4998-55722-0306 & {\hspace{1.15cm}}2.4677$\pm$0.0005 & {\hspace{1.15cm}}... & {\hspace{1.15cm}}no \\
J222559.52$-$004157.5 & {\hspace{1.15cm}}4202-55445-0258 & {\hspace{1.15cm}}2.7721$\pm$0.0004 & {\hspace{1.15cm}}2.776$\pm$0.002 & {\hspace{1.15cm}}no \\ 
J231227.48+005231.7 & {\hspace{1.15cm}}4209-55478-0712 & {\hspace{1.15cm}}3.5710$\pm$0.0011 & {\hspace{1.15cm}}... & {\hspace{1.15cm}}no \\
\enddata
\label{EHVOquasars}
\tablecomments{Information on the 40 EHVO quasars found. QSO column includes the SDSS name, and we have also included the plate-MJD-fiber. The redshifts included are the $z_{PCA}$ provided by SDSS ($z_{\rm em,DR9Q}$), with errors whenever available (see \citealt{Paris12}), and the improved redshift from \citet{Hewett10} whenever available ($z_{\rm em,HW10}$). The BALQSO? column indicates whether the quasar is a {\CIV} BALQSO in the traditional BI definition, defined at lower velocities.}
\end{deluxetable*}

\begin{figure}
\centering
\includegraphics[width=7.7cm]{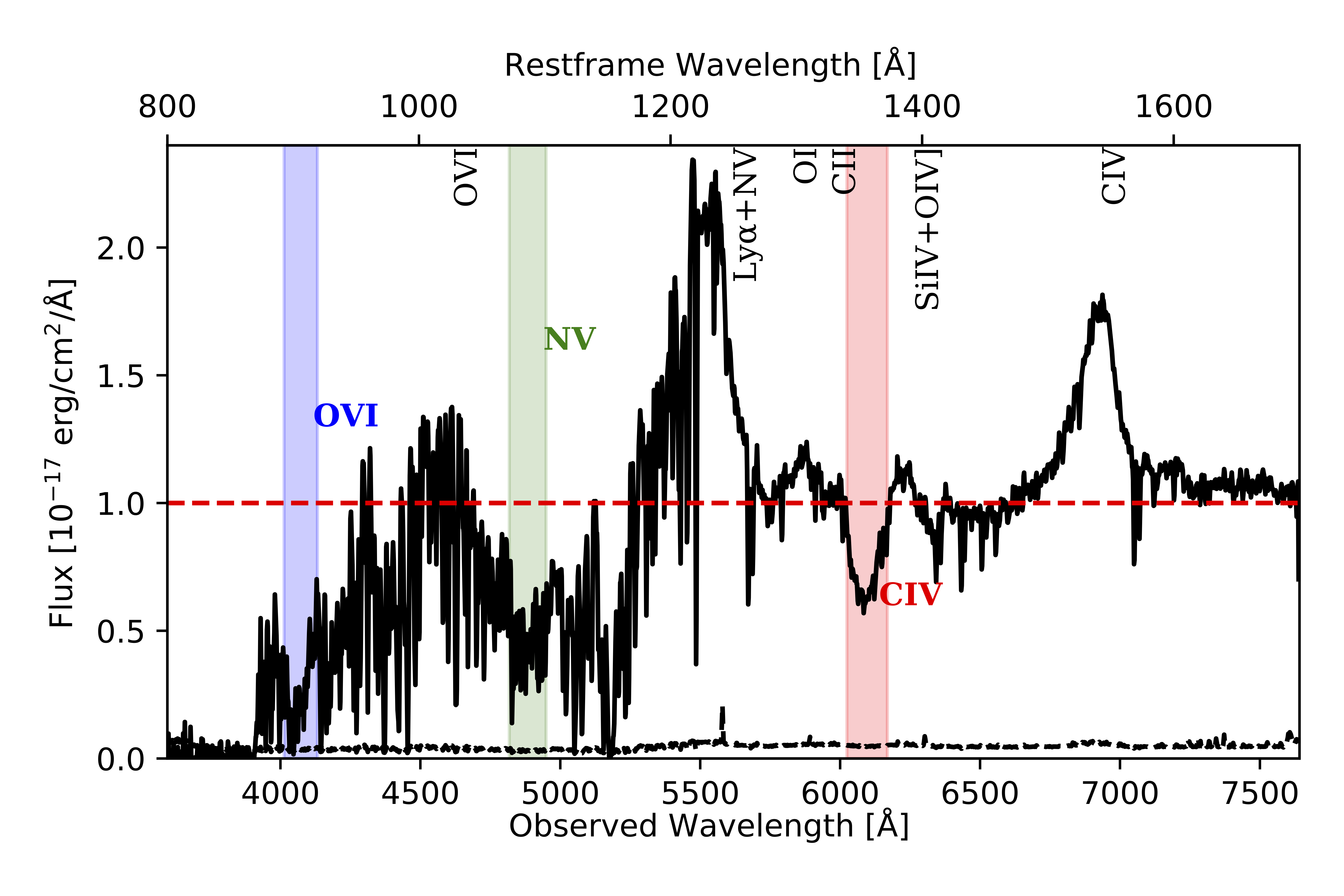}
\includegraphics[width=7.7cm]{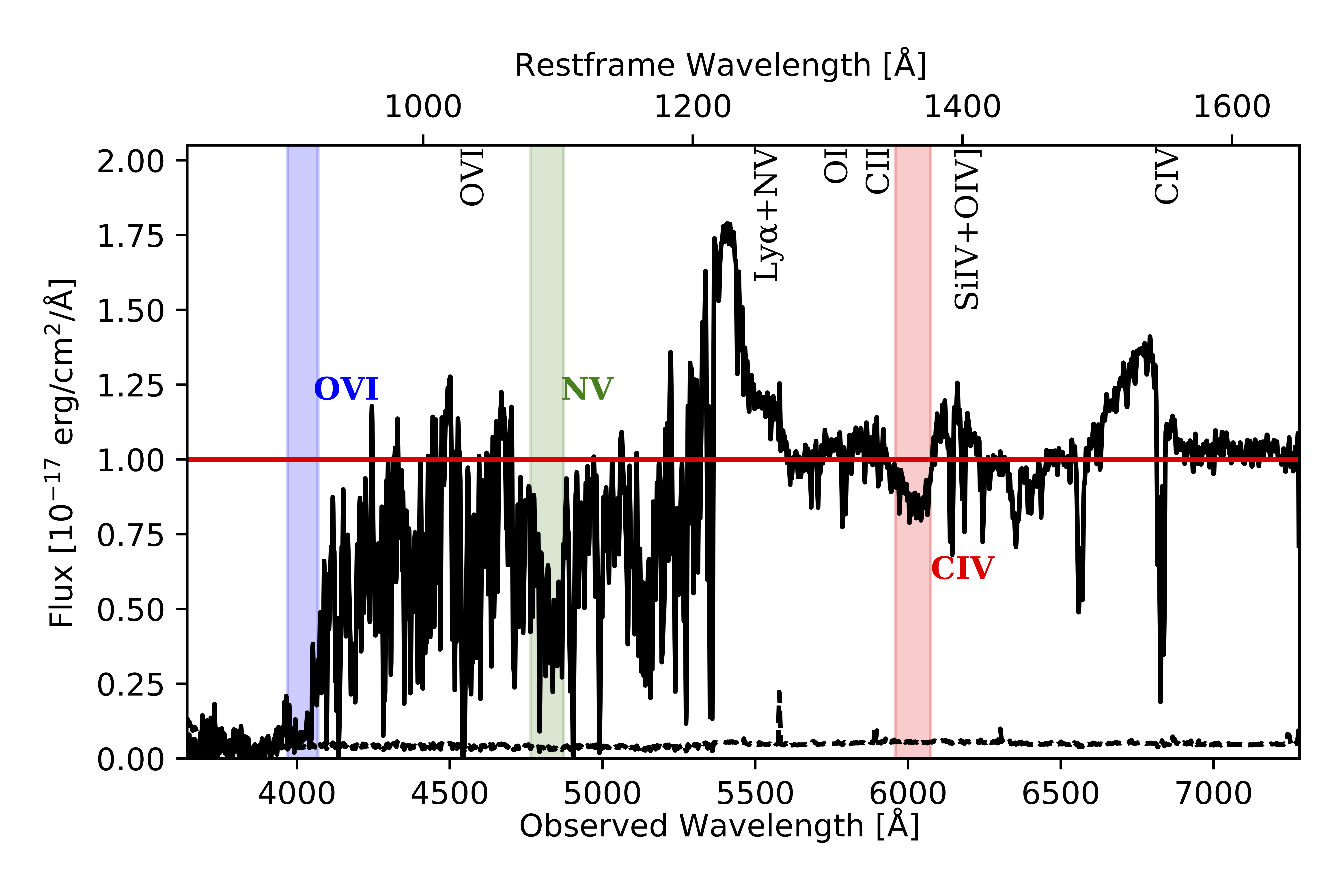}
\includegraphics[width=7.7cm]{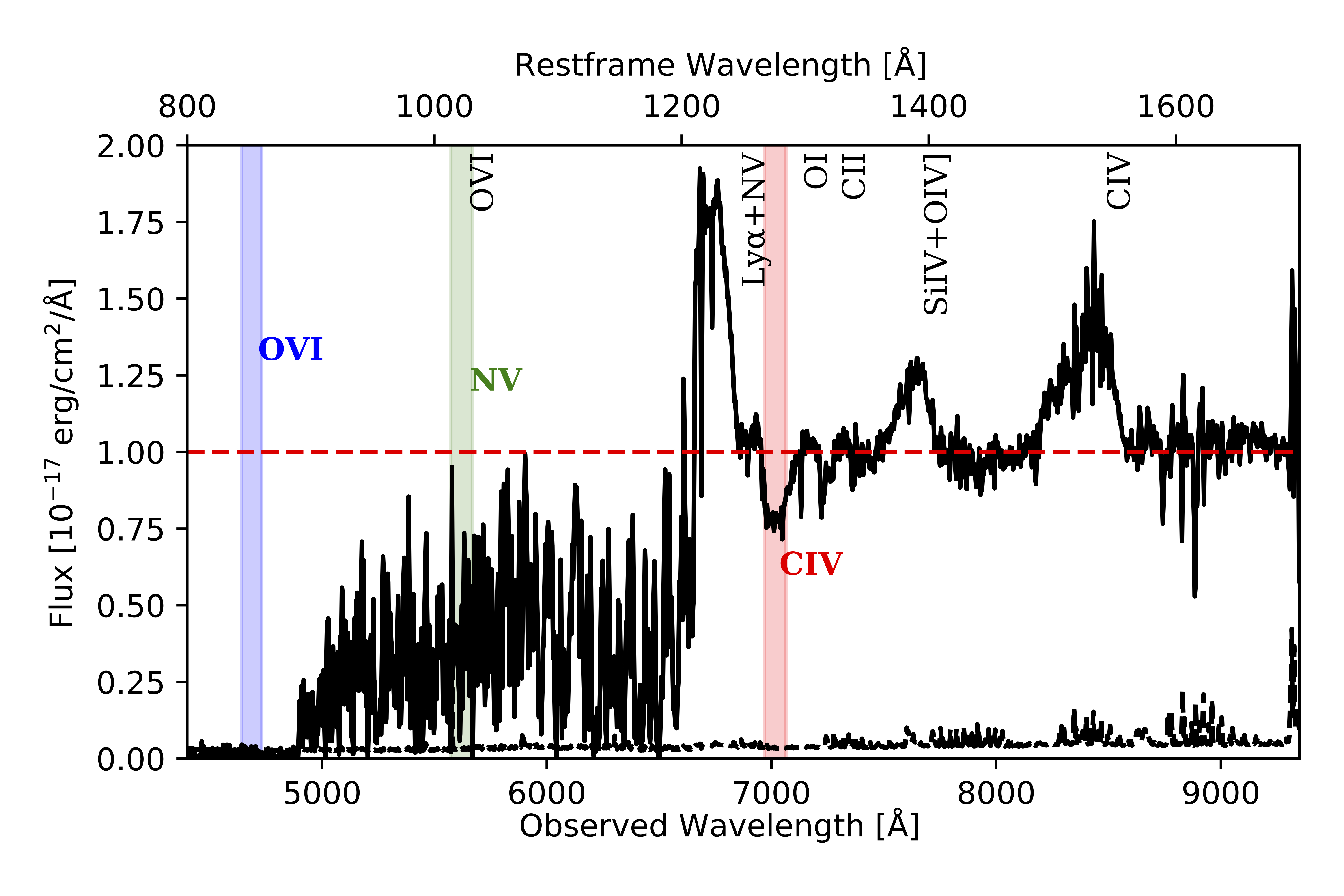}
\caption{Examples of quasars with EHVO {\CIV} absorption, indicated by a red shaded region whose limits are determined by the wavelengths where the absorption crosses the 90\% level of the normalized flux. The green and blue shaded regions are the same lower and upper velocities as the red region but shifted to the corresponding {\NV} and {\OVI} wavelengths, respectively; $v_{max}$ and $v_{min}$ are included in Table \ref{EHVOmeasurements}.
In several cases (top and middle panels), the {\CIV} absorption is clearly accompanied by the corresponding {\NV} outflowing at similar speeds, and in some cases where the wavelength coverage allows, also {\OVI} (top panel). Quasars with very large redshift (bottom panel) show very contaminated Ly$\alpha$ forests, and it is not possible to confirm or reject the presence of {\NV} and {\OVI} absorption. If the absorption was marked as yes, yes? or ? (see Section \ref{sec:other_ions}), we have included shaded regions with the same velocity limits as {\CIV} (red shade) but shifted to the {\NV}, {\OVI}, and Ly$\alpha$ wavelengths (green, blue, and pink shade, respectively). Similar to Figure \ref{SDSSnorm}, some emission lines present in the spectrum are labeled at the top of each figure. (The complete figure set (40 images) is available.)}
\label{fig:EHVOexample}
\end{figure}

\section{Results}
\label{sec:results}

\subsection{Extremely High-velocity Outflow Sample}

We found 40 cases of quasar spectra with EHVO {\CIV} absorption. Table \ref{EHVOquasars} includes some information on these EHVO quasars: QSO name, SDSS DR9Q data collection information (plate, MJD and filter), $z_{\rm em}$ provided by DR9Q ($z_{PCA}$), and whenever available, improved redshifts from \citet{Hewett10} $z_{HW10}$. Finally, we include whether the quasar would be classified as a BALQSO (see below). Figure \ref{fig:EHVOexample} shows several examples where the EHVO {\CIV} absorption is shaded in red. In those cases with sufficiently large redshift, we could search for other ions in the same outflow (for example, {\NV} and {\OVI}, shaded in green and blue in Figure \ref{fig:EHVOexample}, respectively). We discuss the search for other transitions below in Section \ref{sec:other_ions}. 

Seven of the EHVO quasars are BALQSOs in the typical definition based on  absorption present at lower velocities ---  between 5000 and 30,000 {\kms}.
We have indicated this in Table \ref{EHVOquasars} in the ``BALQSO?'' column. However, the BI measurement included in Table \ref{EHVOmeasurements} is the one calculated at extremely high velocities only (see Section \ref{search_abs}). The small number of EHVO quasars in our sample that are also traditional BALQSOs is not necessarily significant: in our goal of compiling a list of secure cases of EHVO quasars, we have rejected 
(or downgraded to potential candidate status)
all of those cases where the potential EHVO absorption could be identified instead as low-velocity {\SiIV} due to the presence of corresponding {\CIV} absorption for a large fraction of the trough (see Figures \ref{fig:rej} and \ref{fig:CandidateExample} in Section \ref{sec:reject}). Some of these cases could have {\CIV} present both at low and high velocities, however, and a large fraction of these would be BALQSOs. 
Thus, we might have systematically removed BALQSOs with potential {\CIV} EHVO absorption. 

\begin{deluxetable*}{ccccccc}
\tablecaption{{\CIV} Absorption Measurements in EHVO Quasars}
\tablewidth{0pt}
\tablehead{
\colhead{QSO} & \colhead{Plate-MJD-fiber} & \colhead{BI$_{\rm EHVO}$} & \colhead{$v_{max}$} & \colhead{$v_{min}$} & \colhead{EW} & \colhead{Depth}\\
 &  & \colhead{(\kms)} & \colhead{(\kms)} & \colhead{(\kms)} & \colhead{(\kms)} & 
}
\startdata
J000154.90$-$004956.4 & 4216-55477-0166 & 200 & -58200 & -54700 & 400 & 0.20\\  
J001306.14+000431.8 & 4217-55478-0020 & 200 & -58200 & -55200 & 200 & 0.31\\ 
J005922.65+000301.4 & 3735-55209-0532 & 200 & -38800 & -35300 & 300 & 0.28\\ 
J012700.69$-$004559.1 & 4229-55501-0337 & 400 & -42100 & -36700 & 400 & 0.33\\
J014548.55$-$000812.5 & 4231-55444-0035 & 1400 & -41500 & -32800 & 1500 & 0.47\\ 
J071843.50+391720.3 & 3655-55240-0388 & 500 & -43500 & -38800 & 700 & 0.26\\ 
J073505.97+280327.0 & 4456-55537-0704 & 400 & -43600 & -37500 & 400 & 0.27\\ 
&  &    & -49500 & -47600 & 50 & 0.19\\ 
J074711.14+273903.3 & 4452-55536-0214 & 1000 & -57900 & -50200 & 1100 & 0.35\\
J075240.18+092523.0 & 4511-55602-0415 & 300 & -42100 & -37800 & 300 & 0.24\\ 
&  &    & -47400 & -45100 & 30 & 0.15\\ 
J075852.68+133530.8 & 4506-55568-0824 & 200 & -37800 & -33700 & 200 & 0.23\\ 
J081337.14+155705.4 & 4498-55615-0502 & 300 & -51200 & -45800 & 400 & 0.26\\ 
J083304.73+415331.3 & 3808-55513-0892 & 2400 & -60000 & -44800 & 2500 & 0.41\\
J085825.71+005006.7 & 3815-55537-0910 & 20 & -38100 & -35800 & 50 & 0.21\\ 
J092125.97$-$023411.9 & 3766-55213-0046 & 500 & -43400 & -36900 & 600 & 0.38\\
J094023.55+404703.2 & 4571-55629-0730 & 900 & -49800 & -41500 & 1000 & 0.33\\
J094258.03+005359.5 & 3826-55563-0860 & 300 & -39800 & -36100 & 500 & 0.40\\ 
J095005.90+362455.2 & 4573-55587-0698 & 500 & -39000 & -34700 & 600 & 0.34\\ 
J095254.10+021932.8 & 4743-55645-0118 & 100 & -36200 & -34000 & 200 & 0.29\\ 
J095603.47+382517.2 & 4570-55623-0296 & 400 & -37500 & -33000 & 500 & 0.34\\ 
J100400.20+372551.9 & 4567-55589-0454 & 800 & -32400 & -30000 & 500 & 0.37\\ 
&  &    & -41900 & -35700 & 500 & 0.31\\ 
J103456.31+035859.4 & 4772-55654-0106 & 40 & -53400 & -51400 & 80 & 0.17\\ 
J110841.93+031735.1 & 4741-55704-0786 & 100 & -40300 & -36300 & 200 & 0.25\\ 
J111154.35+372321.2 & 4622-55629-0678 & 2400 & -51600 & -45700 & 500 & 0.29\\
&  &    & -59900 & -51900 & 2000 & 0.51\\ 
J113000.22+344625.9 & 4619-55599-0854 & 2400 & -42900 & -33800 & 2500 & 0.57\\
J120609.69+004522.6 & 3844-55321-0920 & 1400 & -34000 & -31500 & 300 & 0.28\\
&  &    & -47900 & -38500 & 1200 & 0.32\\ 
&  &    & -56200 & -53700 & 70 & 0.16\\ 
J123056.28+345201.7 & 3968-55590-0464 & 1300 & -42900 & -36000 & 1300 & 0.45\\
J131307.93+065349.0 & 4840-55690-0576 & 700 & -43000 & -37300 & 800 & 0.36\\ 
J133150.25+393417.7 & 4708-55704-0418 & 1300 & -40700 & -32000 & 1400 & 0.50\\
J133752.36$-$011924.7 & 4046-55605-0085 & 200 & -43200 & -39400 & 200 & 0.28\\
J135203.03+333938.3 & 3861-55274-0582 & 400 & -41400 & -36400 & 500 & 0.43\\ 
J144356.21+062539.2 & 4858-55686-0392 & 1100 & -43800 & -35400 & 1200 & 0.33\\
J150339.76+183423.9 & 3957-55664-0568 & 80 & -39500 & -37500 & 90 & 0.24\\ 
&  &    & -48900 & -47700 & 50 & 0.18\\ 
&  &    & -59100 & -57000 & 70 & 0.19\\ 
J151016.40+034200.0 & 4776-55652-0216 & 30 & -48600 & -46500 & 80 & 0.22\\ 
J162445.03+271418.7 & 5006-55706-0846 & 500 & -58400 & -53400 & 500 & 0.32\\ 
J162747.14+192639.7 & 4060-55359-0940 & 60 & -41500 & -39800 & 100 & 0.24\\ 
J164653.72+243942.2 & 4181-55685-0543 & 3600 & -49100 & -36600 & 3700 & 0.71\\
J165436.85+222733.8 & 4178-55653-0608 & 600 & -48400 & -44300 & 600 & 0.39\\ 
J171800.39+314203.9 & 4998-55722-0306 & 1400 & -41200 & -32800 & 1500 & 0.36\\
J222559.52$-$004157.5 & 4202-55445-0258 & 2 & -40000 & -38700 & 10 & 0.14\\ 
J231227.48+005231.7 & 4209-55478-0712 & 300 & -40600 & -37400 & 400 & 0.39\\ 
\enddata
\label{EHVOmeasurements}
\tablecomments{Typical BI and EW errors are $\sim$20\% of their values (approximately hundreds of {\kms}). Typical $v_{min}$ and $v_{max}$ errors are $\sim$200 {\kms}, and depth errors are $\sim\pm$0.02.}
\end{deluxetable*}

Table \ref{EHVOmeasurements} includes absorption information on all of the 40 EHVO quasar spectra. The BI$_{\rm EHVO}$ was calculated per quasar as described in Section \ref{search_abs}. The upper and lower velocity limits for each EHVO, $v_{max}$ and $v_{min}$ respectively, were determined at the wavelengths where the absorption crosses 90\% of the normalized flux level; these are included for each EHVO absorption feature present in the spectrum.
Six of the 40 quasars show two or more absorption troughs in the same spectrum outflowing at speeds larger than 30,000 {\kms}, making a total of 48 separated absorption features in the 40 quasar spectra. We consider those that contribute separately toward the BI value to be distinct absorption features, as explained in Section \ref{search_abs}, and have different values of $v_{min}$ and $v_{max}$.
Equivalent widths (EW) are the integrated values of the absorption between the $v_{min}$ and the $v_{max}$ limits.
Depth measurements were obtained as 1 minus the local minimum flux value of the trough avoiding noise spikes and unrelated narrow absorption; narrow lines contaminating the EHVO trough were masked prior to this calculation. All of our absorption measurements are carried out using the unsmoothed spectra. 
Errors on BI, $v_{max}$, $v_{min}$, EW, and depth are mostly influenced by the location of our continuum fit (a power law, see Section \ref{norm}) since it will shift the location of the normalized flux level; 
we estimated these errors by raising and lowering the normalized continuum by an amount that would place the new continuum fit within the spectrum error 
(typically by 5\% of the normalized flux) and considering the recalculated measurements as 3$\sigma$ deviations. 
The resulting BI and EW $\sigma$ errors are $\sim$20\% of their values -- typically near 100 {\kms} for our BI median value of 400 {\kms} and EW median value of 300 {\kms}.\footnote{Notice that this value is smaller because EWs are included per absorption trough and BI values per total quasar.} Typical $v_{min}$ and $v_{max}$ errors are $\sim$200 {\kms}. Values in Table \ref{EHVOmeasurements} are rounded to reflect significant figures. 

We caution against using either 40/6743 (secure cases) or (40+60)/6743 (secure+candidate cases) as estimates for the {\it percentage} of EHVO quasars among the 6743 parent sample quasar spectra. In this paper, we are presenting a sample of {\it secure} EHVO cases, and we have not included any potentially conflicting cases where the absorption is located on top of the emission or where the absorption might correspond to any other transition. Thus, any estimate derived from the frequency of the cases found in this sample is a very conservative lower limit of the percentage of EHVO cases in quasar spectra.

\subsection{Other possible ionic transitions outflowing at extremely high speeds}
\label{sec:other_ions}

\subsubsection{{\NV} and {\OVI}}
\label{NVOVI}

Our final sample includes 48 EHVO {\CIV} absorption features found in 40 EHVO quasar spectra. Out of those, 26/48 cases have confirmed or likely corresponding {\NV} outflowing at similar speeds, and {\OVI} is likely present in at least 6/48 cases as well. Each of these additional ions confirms that the originally identified absorption is in fact a {\CIV} EHVO, just as \citet{Januzzi96} and \citet{RodriguezHidalgo11} used the presence of {\NV} and {\OVI} in the same outflow to confirm the nature of the first detected cases of {\CIV} EHVOs in quasar spectra.

Table \ref{EHVOotherions} includes our visual identifications of the potential presence of {\NV} and {\OVI} for all the EHVO absorption features.  Detections that are either very likely or likely are marked as `yes' and `yes?,' respectively. Cases where the ionic transition may be present but it is not clearly visually detected are included as `?,' and those where the wavelength coverage did not allow us to determine the presence of {\NV} or {\OVI} are indicated  by `n.c.' for `not covered.'

Figure \ref{fig:EHVOexample} shows several examples of these identifications. The top panel shows a case where both {\NV} and {\OVI} are likely present (`yes' and `yes?' in Table \ref{EHVOotherions}, respectively); while absorption seems to be present at the location of {\OVI}, it lies close to the Lyman edge, and the low flux prevents a secure detection. The middle panel shows a case where {\NV} is detected (`yes') but the presence of {\OVI} is unconfirmed (marked in Table \ref{EHVOotherions} as `?') because it lies too close to a Lyman limit cutoff. The bottom panel shows a case where both {\NV} and {\OVI} are unconfirmed; this was an issue for all the quasars with the largest redshifts in our sample because they have more contaminated Ly$\alpha$ forests where the potential presence of {\NV} and {\OVI} is masked by lower average flux levels.

The presence of {\NV} at similar speeds to the highest EHVO {\CIV} in our sample ($\lvert v_{max} \rvert >55000$ {\kms}) has the additional complication of lying close to the location of the {\OVI}
$\lambda\lambda$1031.9261,1037.6167 emission line, so all those questionable cases appear as `?' in Table \ref{EHVOotherions}. 

We did not carry out a detailed normalization of the Ly$\alpha$ forest. The normalized continuum in this region was derived from extrapolating the power law found at larger wavelengths, without selecting any ``anchor'' wavelength within the Ly$\alpha$ forest. In many cases, especially for the spectra of those quasars with lower emission redshifts, this was sufficient to yield a good fit (see, for example, the middle panel in Figure \ref{fig:EHVOexample}) -- the continuum fit appears to cross the flat, highest-flux levels in this region. However, the combination of emission lines of unknown strength (such as {\OVI}), a myriad of hydrogen absorption lines whose number increases with emission redshift, and the edge of the SDSS spectrum where the sensitivity is reduced makes a correct continuum fit too ambiguous (see, for example, Figure \ref{fig:EHVOexample}: the top spectrum shows an ambiguous case in which the {\OVI} emission line might be stronger and the continuum fit might need to be lower, and the bottom spectrum shows a case with large emission redshift ($z_{\rm em}=$4.478) in which the flux is clearly lower than at redder wavelengths due to abundant hydrogen absorption). 
Additionally, \citet{Telfer02} used a sample of 332 Hubble Space Telescope (HST)/FOS quasar spectra to create a composite spectrum and showed that a broken power law might be required to fit the continuum to the blue of the Ly$\alpha$+{\NV} emission line. 
All of these issues make a correct normalization of the Ly$\alpha$ forest quite difficult and beyond the scope of this paper; thus, we will not provide absorption measurements of any potential ionic transitions at those wavelengths. 

\startlongtable
\begin{deluxetable}{cccc} 
\tablecaption{Other detected transitions in EHVO quasars}
\tablewidth{0pt}
\tablehead{
\colhead{QSO} & \colhead{$v_{max}$} & \colhead{\NV ?} & \colhead{\OVI ?} \\
 &   \colhead{(\kms)} & &
}
\startdata
J000154.90$-$004956.4  & -58200 &  ? & n.c.\\ 
J001306.14+000431.8  & -58200 &  n.c. & n.c.\\ 
J005922.65+000301.4  & -38800 &  ? & ? \\ 
J012700.69$-$004559.1  & -42100 &  yes? & yes? \\ 
J014548.55$-$000812.5  & -41500 &  yes & n.c.\\ 
J071843.50+391720.3  & -43500 & yes & n.c. \\ 
J073505.97+280327.0  & -43600 &  n.c. & n.c. \\ 
                  & -49500 &  n.c. & n.c. \\ 
J074711.14+273903.3  & -57900 &  ? & ? \\ 
J075240.18+092523.0  & -42100 &  yes? & ? \\ 
                & -47400 &  ? & n.c. \\
J075852.68+133530.8  & -37800 &  yes & ? \\ 
J081337.14+155705.4  & -51200 &  ? & n.c.\\ 
J083304.73+415331.3  & -60000 &  n.c. & n.c. \\ 
J085825.71+005006.7  & -38100 &  yes & n.c. \\ 
J092125.97$-$023411.9  & -43400 &  yes & n.c. \\ 
J094023.55+404703.2  & -49800 &  ? & n.c.\\ 
J094258.03+005359.5  & -39800 &  yes & n.c. \\
J095005.90+362455.2  & -39000 &  n.c. & n.c.\\ 
J095254.10+021932.8  & -36200 &  n.c. & n.c.\\ 
J095603.47+382517.2  & -37500 &  yes & ?\\ 
J100400.20+372551.9  & -32400 &  yes & n.c.\\ 
                & -41900 &  yes & n.c.\\
J103456.31+035859.4  & -53400 &  yes & n.c.\\ 
J110841.93+031735.1  & -40300 &  yes & yes?\\ 
J111154.35+372321.2  & -51600 &  n.c. & n.c. \\ 
                & -59900 &  n.c. & n.c. \\ 
J113000.22+344625.9  & -42900 &  yes & yes?\\ 
J120609.69+004522.6  & -34000 &  yes & n.c.\\ 
                & -47900 &  yes & n.c.\\ 
                & -56200 &  ? & n.c. \\ 
J123056.28+345201.7  & -42900 &  yes & yes? \\
J131307.93+065349.0  & -43000 &  yes & n.c.\\ 
J133150.25+393417.7  & -40700 &  n.c. & n.c.\\
J133752.36$-$011924.7  & -43200 &  yes &  n.c.\\ 
J135203.03+333938.3  & -41400 &  yes & yes\\ 
J144356.21+062539.2  & -43800 &  yes & n.c.\\
J150339.76+183423.9  & -39500 &  n.c. & n.c. \\ 
                & -48900 &  n.c. & n.c. \\ 
                & -59100 &  n.c. & n.c. \\ 
J151016.40+034200.0  & -48600 &  n.c. & n.c.\\ 
J162445.03+271418.7  & -58400 &  ? & ?\\ 
J162747.14+192639.7  & -41500 &  ? & n.c.\\ 
J164653.72+243942.2  & -49100 &  yes? & n.c. \\ 
J165436.85+222733.8  & -48400 &  yes? & ? \\ 
J171800.39+314203.9  & -41200 &  yes & n.c.\\ 
J222559.52$-$004157.5  & -40000 &  yes & n.c.\\ 
J231227.48+005231.7  & -40600 &  yes & yes? \\ 
\enddata
\label{EHVOotherions}
\tablecomments{`yes': clear presence of the transition; `yes?': likely presence of the transition; \\
`?': not possible to discern whether the transition is present in the spectrum;  \\
`n.c.' (not covered): the wavelength coverage of the SDSS spectrum does not cover the transition.}
\end{deluxetable}

Visual inspection shows that in all cases where {\NV} is covered by the SDSS wavelength (35) and at least likely to be present (26/35), the depth of the {\NV} absorption feature is at least as deep as the {\CIV} absorption and deeper in many cases (see, for example, the middle panel in Figure \ref{fig:EHVOexample}). Out of the cases with a very likely or confirmed {\NV} detection (`yes' in Table \ref{EHVOotherions} -- 17 cases), nine {\NV} absorption features show similar depths as {\CIV} and eight {\NV} absorption features appear deeper than {\CIV}. These numbers might be, however, biased toward stronger {\NV} that is more easily detected in the {\Lya} forest.

\subsubsection{\rm Ly{$\alpha$}}

We find that strong and broad absorption due to EHVO {\Lya} $\lambda$1215, with a similar absorption profile to the corresponding EHVO {\CIV} trough, to be rare among EHVO features.
We find potential {\Lya} absorption in five EHVO quasar spectra. In four of these cases, broad {\NV} absorption is also present and stronger than the {\Lya} absorption. In one case, the absorption profile of neither {\NV} nor {\Lya} mimics the {\CIV} profile, and so it is not clear whether the absorption is due to {\NV} absorption which is broader than (and perhaps slightly shifted in velocity from) the EHVO {\CIV} absorption, or to {\Lya} which is stronger than {\NV}. In this particular case, the EHVO {\CIV} is the broadest in our sample ($\Delta v \sim$12,500 {\kms}) and the presence of surrounding emission lines might be masking the true velocity profile. In other words, {\CIV} might present strong absorption beyond the detected velocity range if the surrounding emission lines are absorbed by the {\CIV} EHVO. We discuss this interesting case further in Rodr\'iguez Hidalgo et al.\ (2020, in preparation).

Besides the real possibility that Ly$\alpha$ absorption may be weak in EHVO quasars, the small number of spectra with any detected {\Lya} absorption, even weak absorption, might be due to several issues.

First, only strong and broad absorption would be easy to detect among the many other absorption features of the {\Lya} forest; thus, weak absorption might be present but overlooked. Additionally, there is often an overlap between the {\OVI} $\lambda\lambda$1031.9261,1037.6167 emission line and {\Lya} absorption at extremely high velocities. Due to our normalization methodology and the possible ambiguities of a continuum fit to this wavelength region, we will not attempt to provide upper limits to the {\Lya} absorption. 

Second, in cases with very broad {\NV} absorption, absorption due to {\Lya} $\lambda$1215 would lie in the blue wing of {\NV} absorption, which is often seen (see Section \ref{NVOVI}). {\Lya} absorption does not appear to be present and stronger than {\NV}, except for the single potential case mentioned above, because this would create an asymmetric trough stronger at bluer wavelengths that is not observed.

Finally, in cases with several EHVOs in the same spectrum at similar velocities, the {\Lya} corresponding to one {\CIV} absorption feature may appear entangled with {\NV} corresponding to another {\CIV} absorption feature. We do not find in any of those cases stronger absorption at the location of the not-overlapped {\Lya} than at the location of the not-overlapped {\NV}, which would suggest the absorption is due to {\Lya} instead of {\NV}. However, some weak {\Lya} absorption might be contributing to the {\NV} absorption in the overlapped features.

\subsubsection{\SiIV}

At lower velocities, strong {\CIV} absorption is often accompanied by (typically weaker) {\SiIV} absorption. In EHVO quasars, unfortunately, the {\SiIV} absorption tends to lie on top of the {\Lya+\NV} emission-line complex. This blend of emission lines has a diverse range of relative strengths and widths, and a correct fit typically requires including multiple components. Due to the fact that we searched only for absorption below the continuum level, without fitting a pseudo-continuum including emission lines, we cannot search for EHVO {\SiIV} absorption in those cases. In other cases, the EHVO {\SiIV} would be present in the {\Lya} forest. However, by visual inspection, we did not find any case where broad EHVO {\SiIV} that mimics the {\CIV} absorption profile is clearly present.

\subsection{Analysis of {\HeII} Emission lines in EHVO quasars}
\label{sec4.3}

Photoionization is the most important factor in determining quasar broad emission-line fluxes.  Photoionization should affect the strength of broad absorption lines as well as broad emission lines. Although broad emission- and absorption-line strengths should be correlated at some level, those correlations will be smeared due to light travel time effects on emission lines and density-dependent response times of absorption lines. 

\begin{figure}
\centering
\includegraphics[width=8cm]{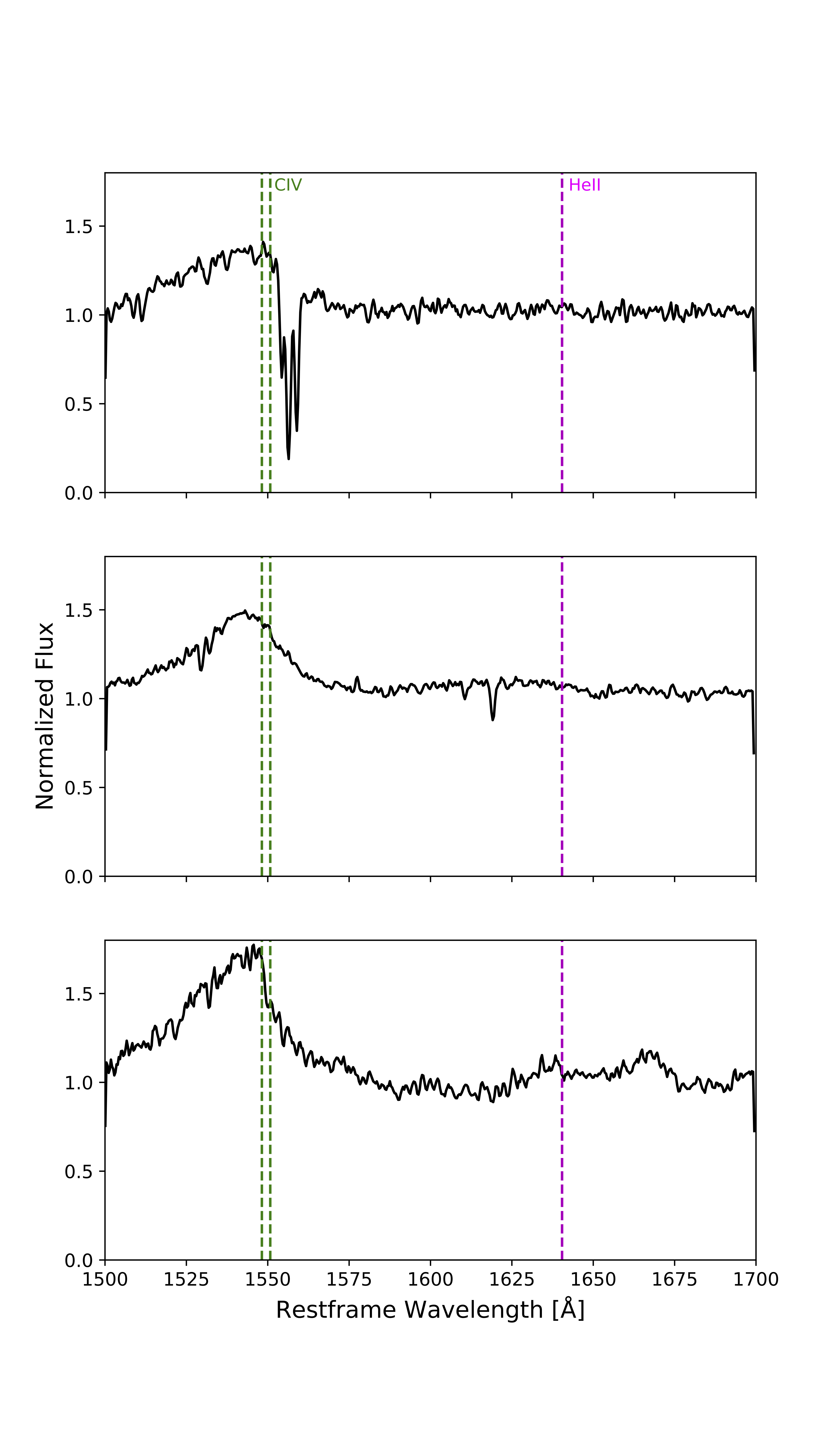}
\caption{Examples of wavelength region where {\HeII} should be present in our sample of EHVO quasar spectra. The {\HeII} location at the adopted quasar redshift is indicated by a magenta dashed line, and the location of the {\CIV} emission doublet line is marked in green. 
Most of the cases in our sample are similar to the top panel, in which the region where {\HeII} would be present is practically flat. Approximately 20\% of our sample shows some broader emission that cannot be unambiguously attributed to {\HeII} (middle panel). Only three cases clearly show the presence of {\HeII} emission, (bottom panel) together with {\OIII}$\lambda$1660.8,1666.2. Emission-line peaks appear blueshifted, as expected. 
}
\label{fig:HeIIEHVOexample}
\end{figure}

The strength of the {\HeII} $\lambda$1640.42 emission line is known to be related to the presence of soft X-ray continuum emission. \citet{Casebeer06} discussed the dependence between different spectra energy distributions (SEDs) and the strength of the different emission lines including {\HeII}. They showed that the {\HeII} emission line is stronger for harder SEDs---in other words, when a larger fraction of higher-energy photons is emitted. In those conditions, observing strong winds is less likely because the gas may be too ionized to produce observable absorption. Indeed, \citet{Richards11} showed that BAL-type quasars have typically weaker {\HeII} emission in their composite spectra (their Figures 11 and 12) and predicted that quasars with strong {\HeII} emission will be less likely to show outflows as absorption (see also \citealt{Baskin15}). 

We investigated the relative strength of the {\HeII} $\lambda$1640.42 emission line in all our EHVO quasars. We follow a similar procedure to \citet{Baskin13} and \citet{Baskin15} to estimate the {\HeII} EW by integrating the normalized flux in the $\lambda_{\rm rest}$ range between 1620 and 1650 {\AA}. The Baskin et al.\ normalizing window (1700-1720 {\AA}) overlaps with our normalization region A.
Nonetheless, we divided each spectrum by the mean in that range again for an accurate comparison. We obtained a mean EW$_{\HeII}$ of 0.505 {\AA}. 
However, our quasars' spectra might have contamination at the red edge of the 1700-1720 {\AA} range due to blueshifted emission lines. We decided to use both regions A and B (1677 - 1725 {\AA}) as our normalizing window. 
With that normalization, we obtained a mean EW$_{\HeII}$ of 0.83 {\AA}, which is most similar to the lowest quartile value of 0.9 {\AA} for low-$z$ BAL quasars in \citet{Baskin13} (see their Table 2) and to the value of 0.4 {\AA} for the low-EW bin of high-$z$ BAL quasars (see their Table 1).
None of our values are larger than 5.4 {\AA}, which was the median of their high-EW bin.

When measuring EW$_{\HeII}$, we might be integrating {\FeII} emission lines that tend to appear as a plateau redward of the {\CIV} emission line. Weak {\HeII} emission line would be indistinguishable from this {\FeII} emission, but relatively strong {\HeII} would show a peak at the corresponding wavelength, such as is observed in Figure 5 of \citet{Baskin13}, which would confirm the presence of {\HeII}. 
We searched for this peak by measuring the mean flux in the 1630--1650 {\AA} wavelength range as compared to the mean flux in the 1605--1625 {\AA} range, combined with visual inspection of the spectra.

In Figure \ref{fig:HeIIEHVOexample}, we show examples of different strengths of potential {\HeII} found in our analysis. In our sample, most cases show no indication of emission of {\HeII} in that region (top panel), as both the EW and the relative flux measurements confirm. In $\sim$20\% (7/40) of cases, we see some broad emission, confirmed by the {\HeII} EW $>$ 2{\AA}, but no peak at 
$\sim$1640 {\AA} is observed, so it might be more likely to be {\FeII} emission (middle panel). 
Only in three cases do we find an emission peak at the expected location of the {\HeII} emission line (bottom panel in Figure\ref{fig:HeIIEHVOexample}), confirmed by either the EW or the relative mean flux. Notice that the {\OIII]}
$\lambda$1660.8,1666.2 emission lines are also present in this case.

In conclusion, our results suggest
that {\HeII} is very weak in EHVO quasars, even weaker on average than in BAL quasars in general.

\subsection{Bolometric Luminosities, Black Hole Masses, and Eddington Ratios}
\label{sec4}

\begin{deluxetable*}{cccc}
\tablecaption{DR7 Properties of DR9 EHVO quasars \label{Table4}}
\tablehead{
\colhead{EHVO Quasar} & \colhead{log($L_{\rm bol}$/erg~s$^{-1}$)} & \colhead{log($M_{\rm BH}$/$M_{\odot}$)} & \colhead{log($L_{\rm bol}$/$L_{\rm Edd})$}
}
\startdata
001306.14+000431.8 & 47.065$\pm$0.010 & 9.80$\pm$0.06 & -0.83 \\
005922.65+000301.4 & 47.29$\pm$0.03 & 9.43$\pm$0.07 & -0.35 \\
012700.69$-$004559.1 & 47.584$\pm$0.014 & 9.85$\pm$0.05 & -0.37 \\
014548.55$-$000812.5 & 46.853$\pm$0.015 & 9.88$\pm$0.09 & -1.21 \\  
074711.14+273903.3 & 47.750$\pm$0.011 & 10.46$\pm$0.15 & -0.81 \\
075852.68+133530.8 & 47.421$\pm$0.010 & 10.02$\pm$0.06 & -0.70 \\
083304.73+415331.3 & 46.740$\pm$0.012 & 9.1$\pm$0.5 & -0.47 \\
085825.71+005006.7 & 47.146$\pm$0.012 & 9.30$\pm$0.11 & -0.17 \\
095005.90+362455.2 & 46.958$\pm$0.007 & 9.57$\pm$0.08 & -0.72 \\
095254.10+021932.8 & 47.153$\pm$0.006 & 9.56$\pm$0.05 & -0.51 \\
103456.31+035859.4 & 47.625$\pm$0.005 & 9.88$\pm$0.05 & -0.37 \\
111154.35+372321.2 & 47.259$\pm$0.005 & 10.10$\pm$0.06 & -0.94 \\  
113000.22+344625.9 & 47.316$\pm$0.011 & 9.03$\pm$0.08 & 0.24 \\
123056.28+345201.7 & 47.470$\pm$0.007 & 9.82$\pm$0.04 & -0.53 \\
133150.25+393417.7 & 46.217$\pm$0.019 & 9.76$\pm$0.10 & -1.65 \\   
133752.36$-$011924.7 & 46.68$\pm$0.02 & 9.47$\pm$0.13 & -1.31 \\  
135203.03+333938.3 & 46.87$\pm$0.03 & 9.9$\pm$0.5 & -1.15 \\
162445.03+271418.7 & 47.540$\pm$0.012 & 10.13$\pm$0.17 & -0.69 \\
164653.72+243942.2 & 47.221$\pm$0.016 & 10.02$\pm$0.07 & -0.90 \\
165436.85+222733.8 & 47.745$\pm$0.013 & 10.1$\pm$0.6 & -0.55 \\
222559.52$-$004157.5 & 47.548$\pm$0.007 & 9.50$\pm$0.03 & -0.02 \\
\hline 
Median,std & 47.3,0.4 & 9.8,0.4 & -0.7,0.4 \\
\enddata
\end{deluxetable*} 

In this section, we study the parameter space of physical properties of the EHVO quasars in our sample versus their parent sample to learn whether they represent a special subgroup of quasars with particular properties. 
While DR9Q does not include measurements of bolometric luminosity ($L_{\rm bol}$), black hole mass ($M_{\rm BH}$) or Eddington ratio ($L_{\rm bol}/L_{\rm Edd}$), those values have been published for quasars in the SDSS Data Release 7 (DR7) in \citet{Shen11}. We cross-correlated our final sample with this catalog, which resulted on 21 EHVO quasars  present in both DR7 and DR9. Table \ref{Table4} includes the values directly from \citet{Shen11} or derived from their data (see explanation below). The last line of Table \ref{Table4} shows the median and dispersion values for each property. 

We used the value and error of $L_{\rm bol}$ provided by \citet{Shen11}. For quasars with $z_{\rm em}\geq 1.9$, they calculated $L_{\rm bol}$ using the value of $L_{1350}$ from their spectral fits and a bolometric correction BC$_{1350}$=3.81 from the composite spectral energy distribution in \citet{Richards06}. Bolometric corrections were carried out for all DR7 quasars similarly, independently of whether or not the quasar was a BALQSO. No correction for intrinsic reddening was applied. BALQSOs are known to be redder than non-BALQSOs (see, e.g., \citealt{Reichard03}; 
\citealt{Gibson09b}).
Additionally, using $L_{1350}$ to calculate  $L_{\rm bol}$ is a potential issue as absorption might be present at that restframe wavelength for both BALQSOs and EHVO quasars. 
Thus, the true $L_{\rm bol}$ might be larger than the calculated $L_{\rm bol}$ from \citet{Shen11}. 
In fact, \citet{Shen11} cautioned against the use of individual values of $L_{\rm bol}$. However, we are using $L_{\rm bol}$, as recommended, to study the average properties of these samples.
We discuss in Section \ref{sec:4.4.1} how these issues might be affecting our findings, and in Section \ref{sec4.5} we show the analysis of $M_i[z=2]$ that supports these results.

Our values and errors for $M_{\rm BH}$ in Table \ref{Table4} are the adopted fiducial virial black hole mass in \citet{Shen11}. 
For quasars with $z_{\rm em}\geq$1.9, \citet{Shen11} used the {\CIV} estimates from \citet{Vestergaard06}, which have been found to be somewhat overestimated at times, especially for those cases with large {\CIV} blueshifts that might correspond to material that is not virialized (\citealt{Shen08}). Quasars with large luminosities and BALQSOs are known to have particularly large {\CIV} blueshifts (e.g., \citealt{Paris12}). 
Given that EHVO quasars overall show large values of $L_{\rm bol}$ relative to DR9Q (see Figure \ref{Lbol}), EHVO quasars likely have overestimated $M_{\rm BH}$ values as well.
While we do not know the {\CIV} blueshift values for all of our EHVO quasars, we calculated a potential correction described in  \citet[their equations (4) and (6)]{Coatman17}.
We lack sufficient information on the {\CIV} blueshifts to apply this correction to our sample of EHVOs, but for those cases for which we have {\CIV} blueshifts (10 out of the 21 cases; from Allen \& Hewett 2020, in preparation), we find that \citeapos{Coatman17} correction for $\log(M_{BH}/M_{\odot})$ would have been only 
$-$0.08, resulting in smaller $M_{\rm BH}$ values. 
We did not apply this correction to our measurements. 

Eddington ratios (log\,$L_{\rm bol}/L_{\rm Edd}$) were taken from \citet{Shen11}, who used the fiducial virial black hole mass to calculate them.

Throughout this section, we assume that the DR7 information is valid for EHVO quasars detected in DR9. 
We expect at most minor changes in the properties we study, 
as more luminous quasars tend to vary less than lower luminosity quasars \citep{Kaspi07}. 
We defer the study of the variability of these quasars to a future paper.

\subsubsection{Comparison to Parent Sample and BALQSOs}
\label{sec:4.4.1}

To compare the parameter space of properties described above for the 21 EHVO quasars with DR7 information and other classes of quasars, we cross-correlated the parent sample of quasars (6743) with the DR7 database of \citet{Shen11} and found 2883 quasars in the parent sample with DR7 information. 

In section \ref{sec:data}, we describe the cutoffs we performed over the original sample to obtain our parent sample. These cutoffs already select the brightest quasars within DR9Q due to the S/N constraint. Therefore, we are not comparing our sample to DR9Q as a whole. Notice as well that because we only include those DR9Q quasars in the parent sample with information in \citet{Shen11}, we are selecting a subsample of the parent sample that had previous observations in DR7.

To compare EHVO quasars to BALQSOs (quasars with strong and wide absorption at lower velocities), we selected cases within the parent sample that had been flagged as having BALs in \citet{Paris12}. We constrained the BAL identification by selecting quasars that include all of the following requirements: (1) BAL flag from visual inspection, (2) absorption index AI $>$0, and (3) $|v_{min}| >$ 5000.

We used constraint (1) because the \citet{Paris12} BAL flags from visual inspection are robust at high S/N. While the authors caution against blind uses of the catalog at low S/N, this is not an issue in this case given that our parent sample has already a high S/N cutoff. 
To compare similar widths of absorption in both samples, we used constraint (2). Our definition of BI is set for a width of 1000 {\kms}, similar to the AI definition in \citet{Paris12} and, originally, in \citet{Hall02}. 
Constraint (3) was used to remove possible contamination of ``associated'' absorption from our comparison between BALQSOs and EHVO quasars. \citet{Knigge08} found a bimodal distribution of quasars because the AI definition, when applied from zero velocity, includes all absorption at redshifts close to the quasars' emission redshift. This so-called associated absorption might not be located near the central engine of the quasar but in the host galaxy or in neighboring galaxies, and their low-velocity absorption might represent a completely different type of phenomenon that is difficult to distinguish from low-velocity outflows. After applying all of these cutoffs, we found 444 BALQSOs in the DR9 parent sample; after cross-correlation with DR7, we found 185 BALQSOs. Three of the EHVO quasars are also BALQSOs.

\begin{figure}
\centering
\includegraphics[scale=0.6]{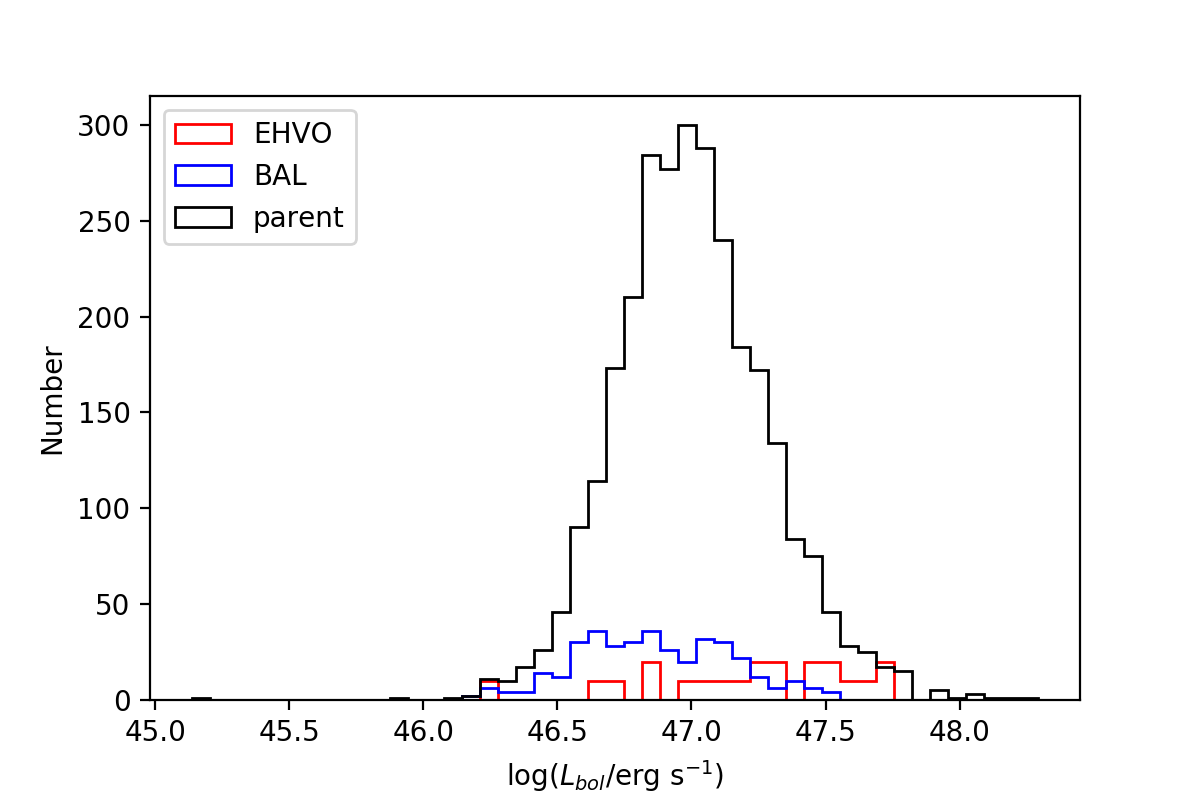}
\caption{Distribution of values of bolometric luminosity for the parent sample of quasars with information in DR7 (2883 quasars -- parent in black), quasars with at least one BAL with width larger than 1000 {\kms} and outflowing with $v_{min} >$ 5000 {\kms} (185---BAL in blue), and quasars with at least one BAL with width larger than 1000 {\kms} at $v_{min} >$ 30,000 {\kms} (21---EHVO in red). The frequencies for EHVO and BAL quasars in each bin have been multiplied by 10 and 2, respectively, to include them in the same scale.}
\label{Lbol}
\end{figure}

\begin{figure}
\centering
\includegraphics[scale=0.6]{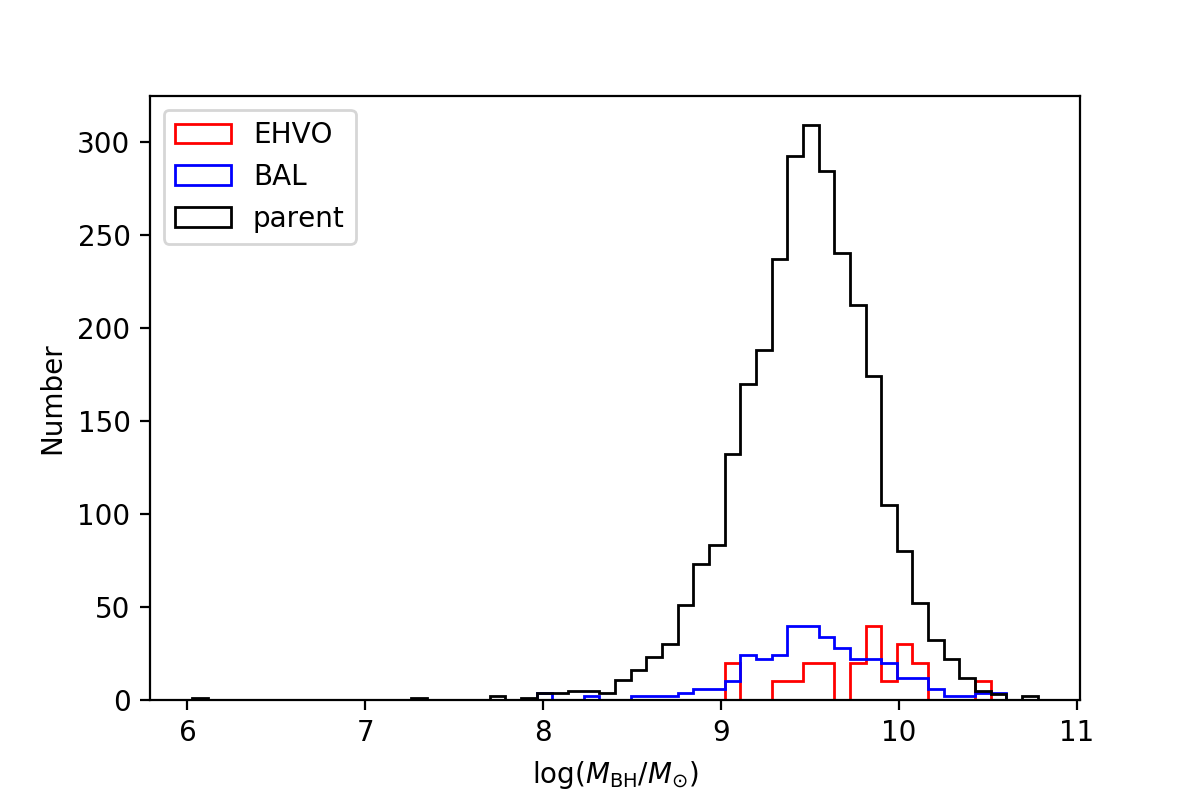}
\caption{Same as Figure \ref{Lbol} for $M_{\rm BH}$ values.}
\label{MBH}
\end{figure}

\begin{figure}
\centering
\includegraphics[scale=0.6]{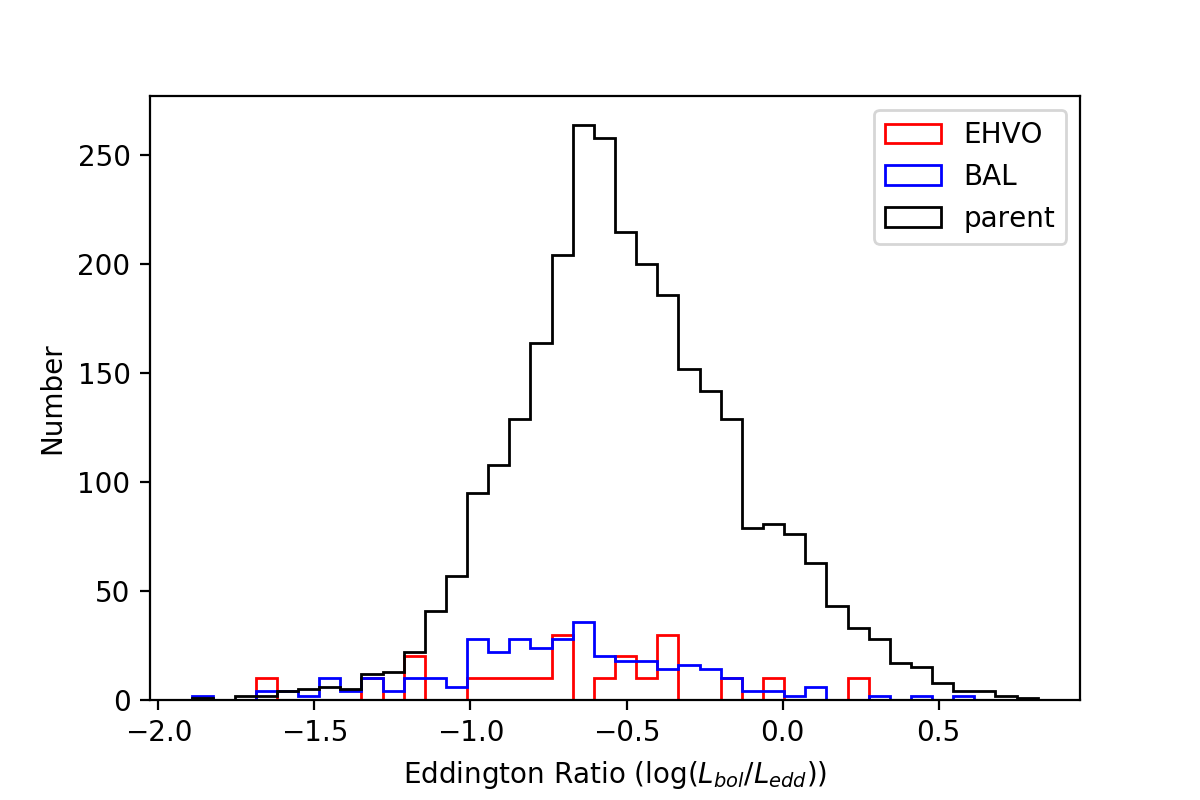}
\caption{Same as Figure \ref{Lbol} for values of Eddington ratios.}
\label{Ledd}
\end{figure}

Figures \ref{Lbol}--\ref{Ledd} show how the properties of the 21 DR7+DR9Q EHVO quasars compare to both the parent sample (2883 cases) and the BAL sample (185 cases) from DR7+DR9Q.
Due to the smaller number of EHVO and BAL quasars, we have multiplied their frequencies in each bin by 10 and 2, respectively, so their distributions can be compared on the same scale.

\begin{deluxetable*}{cccc}
\tablecaption{Median Values of the DR7 properties for the Parent, BALQSO, and EHVO Quasar Samples \label{Table5}}
\tablewidth{0pt}
\tablehead{
\colhead{SDSS DR9 Quasar Name} & \colhead{Median log($L_{\rm bol}$/\text{erg s$^{-1}$})} & \colhead{Median log($M_{\rm BH}$/$M_{\odot}$)} & \colhead{Median log($L_{\rm bol}$/$L_{\rm Edd})$} 
}
\startdata
DR9Q parent sample  & 47.0 & 9.5 & -0.5 \\
BALQSOs  & 46.9 & 9.5 & -0.7 \\
EHVOs  & 47.3 & 9.8 & -0.7 \\
\enddata
\label{Table5} 
\end{deluxetable*} 

EHVO quasars show larger values of $L_{\rm bol}$ and $M_{\rm BH}$ than their parent sample and BALQSOs. This can be observed in Figure \ref{Lbol} and \ref{MBH}, as well as in Table 5, where we show the median values of the three parameters for the parent sample, and the BALQSO and EHVO subsamples. 
The median value of $L_{\rm bol}$ in EHVO quasars exceeds both the median value in the parent sample and the median of the subsample of BALQSOs.
The median of $L_{\rm bol}/L_{\rm Edd}$ is similar for BALQSOs and EHVOs and the distributions overlap significantly (see Figure \ref{Ledd}).

We performed statistical tests to determine the likelihood of obtaining these distributions of EHVO properties randomly from the parent and BALQSO samples. Specifically, we carried out Kolmogorov–Smirnov (K-S) tests between the parent and EHVO sample and between the BALQSO and EHVO distributions of $L_{\rm bol}$, $M_{\rm BH}$ and $L_{\rm Edd}$. 

\begin{table}
\begin{center}
\caption{Results of the Statistical Analysis between EHVO Quasars, BALQSOs, and the parent sample for values of $L_{\rm bol}$, $M_{\rm BH}$, and $L_{\rm bol}/L_{\rm Edd}$}
\begin{tabular}{cccc}
\hline
& &  $p$-values  \\
\hline
 Sample & $L_{\rm bol}$ & $M_{\rm BH}$ & $L_{\rm bol}/L_{\rm Edd}$   \\
\hline
EHVO - parent sample  & 2.7e-3 & 1.1e-3 & 0.18 \\
EHVO - BALQSOs  & 8.8e-05 & 9.3e-03 & 0.70 \\
\hline
\label{Table6} 
\end{tabular} 
\end{center} 
\end{table}

The results are given in Table \ref{Table6}.
We find that EHVO quasars are statistically different from the parent sample and from BALQSOs in their distributions of $L_{\rm bol}$ and $M_{\rm BH}$, but not in the distributions of Eddington ratios.  
It is especially relevant that we find such a small probability that the BALQSO and EHVO samples are derived from the same populations. 
While we find that our $M_{\rm BH}$ values might be slightly overestimated (see Section \ref{sec4}), our $M_{\rm BH}$ values for BALQSOs will be shifted in the same direction because they are also found to have large {\CIV} blueshifts (e.g., \citealt{Richards11}), making our relative comparison valid. 
We also note that the differences in $M_{\rm BH}$ between EHVOs and the parent or BALQSO samples are larger than can be explained purely by the partial dependence of $M_{\rm BH}$ (as calculated by \citealt{Shen11}) on $L_{\rm bol}$.

The small differences found in $L_{\rm bol}$ are affected by how $L_{\rm bol}$ was determined, but little is known about the intrinsic properties of EHVO quasars, so we explore potential shifts of the $L_{\rm bol}$ distributions due to reddening and differential bolometric corrections. 
Reduction of $L_{1350}$ by reddening of a quasar's spectrum by nuclear or host galaxy dust, or by broad absorption, would shift the $L_{\rm bol}$ calculated by \citet{Shen11} to smaller values than the true $L_{\rm bol}$ value.
Such a shift is likely seen in Figure \ref{Lbol} for the BALQSO sample, which is expected because BALQSOs are known to be redder than normal quasars \citep{Reichard03}.
If EHVO quasars have the same reddening distribution as BALQSOs, then EHVOs tend to have larger bolometric luminosities than BALQSOs.
Alternatively, for reddening to explain the $L_{\rm bol}$ distribution for the EHVO sample, EHVO quasars would have to be less reddened than normal quasars.
We have no ready explanation for why EHVO quasars would be less reddened than normal quasars, but
we note that it might be natural for EHVOs to exhibit less reddening than BALQSOs if EHVOs are preferentially seen along high-latitude sight lines. Models that reproduce BALQSOs find outflowing gas along low-latitude sight lines that are more likely to intersect lines of sight through a dusty wind or torus (see, for example, \citealt{Proga04}), while faster outflows are found at higher latitudes.
Similarly, if absorption in EHVO quasars is reducing the value of $L_{1350}$,
it would be shifting the calculated $L_{\rm bol}$ for EHVO quasar to values smaller than the true $L_{\rm bol}$ value. Therefore, our finding that they show large values of $L_{\rm bol}$ instead is even more significant. In Section \ref{sec4.5}, we present an analysis of $M_i[z=2]$ that supports these results.

The possibility of differential bolometric corrections between EHVOs, BALQSOs, and non-BAL quasars cannot be ruled out from SDSS data alone.  However, the fact that as a population BALQSOs have weaker \HeII\ emission than non-BAL quasars, and that EHVOs have weaker \HeII\ emission than BALQSOs (see Section \ref{sec4.3}), suggests that BALQSOs have weaker far-UV emission than non-BAL quasars, and that EHVOs have weaker far-UV emission than BALQSOs.  Therefore, a differential bolometric correction relative to non-BAL quasars arising from different far-UV SEDs would be expected to have the same sign for BALQSOs and EHVOs. Smaller bolometric corrections could reconcile EHVO quasars with the parent population; however, a correction of the same sign will not reconcile both BALQSOs and EHVOs with the general quasar population (see Figure \ref{Lbol}).  Therefore, a differential bolometric correction is unlikely to be the sole explanation for the different $L_{\rm bol}$ distributions of both EHVOs and BALQSOS from non-BAL quasars. More work is needed to better determine the intrinsic properties of EHVO quasars.

\begin{figure}
\centering
\includegraphics[scale=0.55]{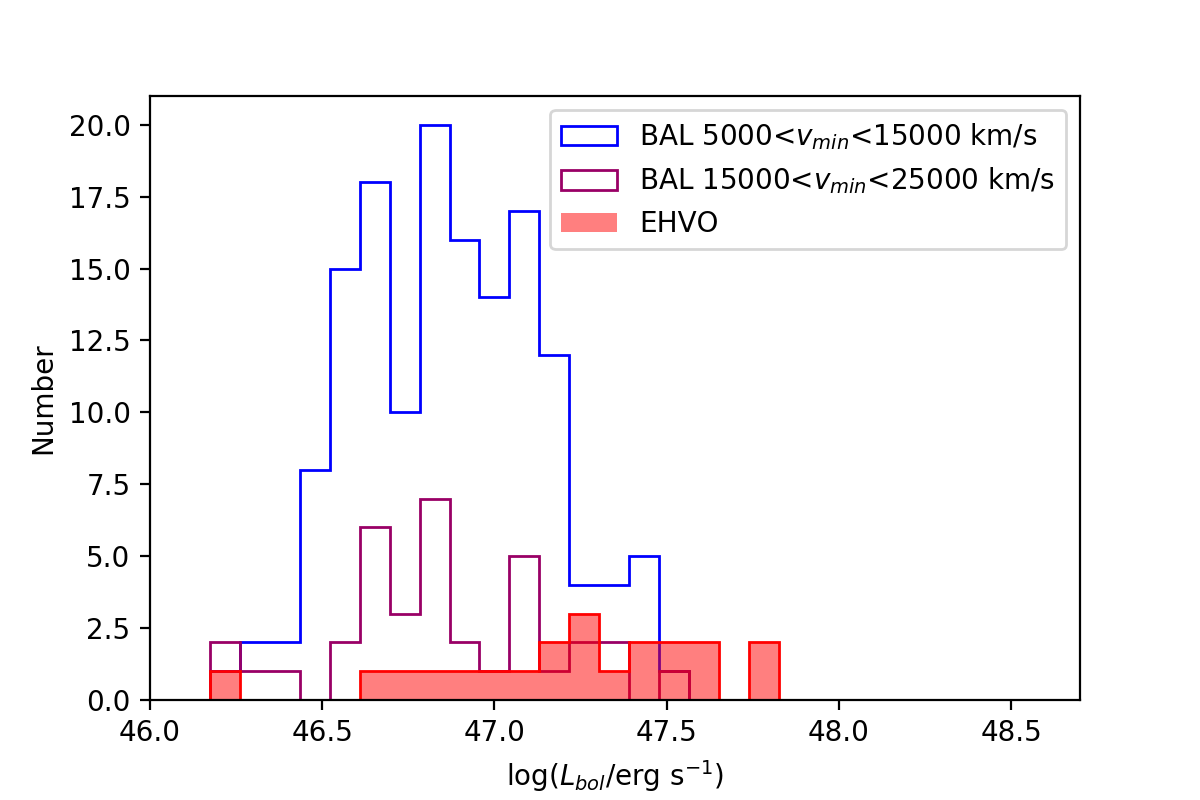}
\includegraphics[scale=0.55]{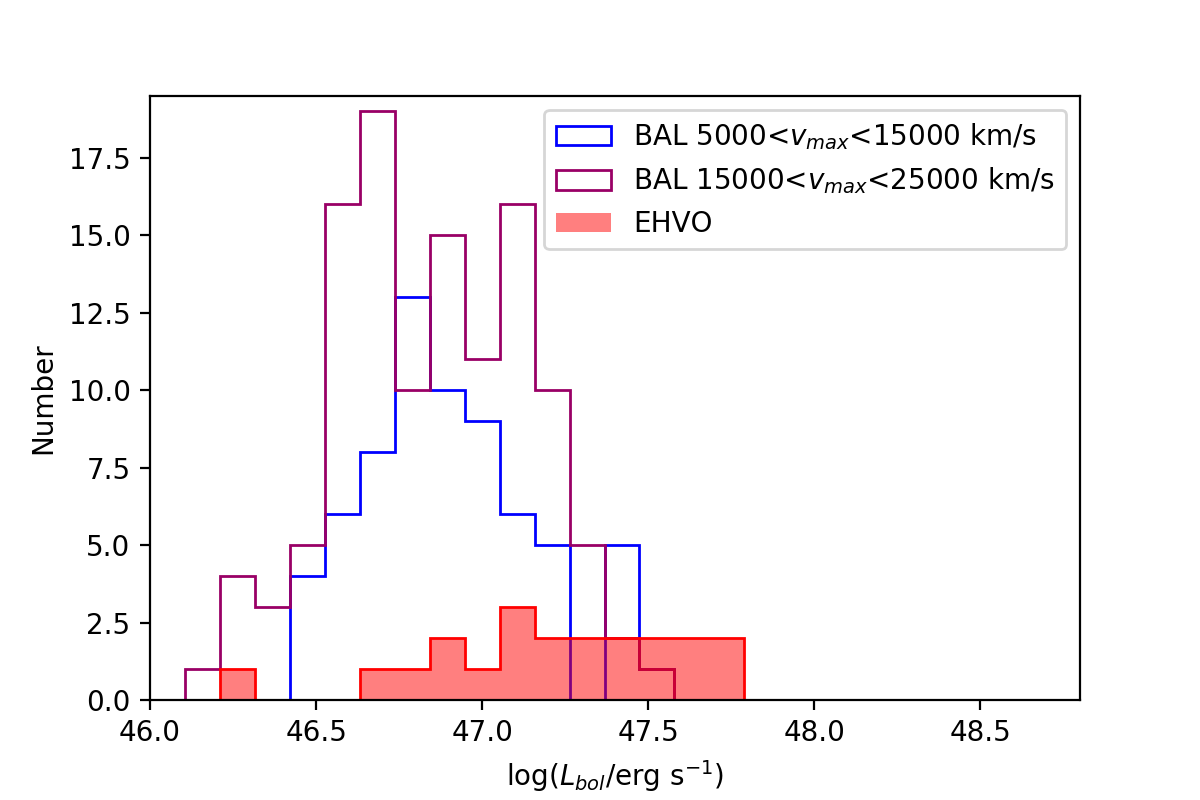}
\caption{Top: 
distribution of $\log(L_{\rm bol})$ values for quasars with at least one BAL with width larger than 1000 {\kms} and outflowing with 5000 $< v_{min} <$ 15,000 {\kms} (BAL in blue), quasars with at least one BAL with width larger than 1000 {\kms} and outflowing with 15,000 $< v_{min} <$ 25,000 {\kms} (BAL in burgundy), and quasars with at least one BAL with width larger than 1000 {\kms} at $v_{min} >$ 30,000 {\kms} (EHVO in red). The frequencies for EHVO quasars in each bin have not been multiplied by any factor. 
The population of EHVO quasars shows larger $L_{\rm bol}$ values and no progression toward larger $L_{\rm bol}$ is observed as the $v_{min}$ of BALQSOs increases. Bottom: similar to the top figure, but displaying $v_{max}$ instead of $v_{min}$. No progression in BALQSOs $v_{max}$ is observed either.}
\label{Lbol_BAL}
\end{figure}

BALQSOs and EHVO quasars appear to show distinct $L_{\rm bol}$ and $M_{\rm BH}$ properties. Because the major observational differences in their spectra are the minimum and maximum velocities of their {\CIV} outflows, we investigated the relation between their outflow velocities and these properties. In Figure \ref{Lbol_BAL}, we show the distribution of $L_{\rm bol}$ for EHVO and BALQSOs, where the BALQSOs have been divided into groups based on the minimum and maximum velocities of their {\CIV} outflows. These velocities were extracted from \citet{Paris12}.\footnote{Note that \citet{Paris12} use $v_{min}$ and $v_{max}$ in the opposite sense from we do.} No progression toward higher $L_{\rm bol}$ is observed with increasing $v_{min}$ in BALQSOs; all the median values of $\log(L_{bol})$ are within 0.07 of 46.83, while the median for EHVO quasars is 47.26, and all $p$-values between subsets of values of $\log(L_{\rm bol})$ for BALQSOs are larger than 0.5. A similar result is observed when selecting BALQSOs based on their $v_{max}$: all of their median values are within 0.02 of 46.88. We find the $p$-values between subsets selected based on $v_{max}$ show smaller values overall than for $v_{min}$, but none of them are smaller than 0.03; we cannot conclude that those subpopulations of BALQSOs are distinct. On the other hand, all comparisons in $\log(L_{\rm bol})$ between the EHVO population and subsets of BALQSOs larger than the EHVO sample have $p$-values are smaller than 0.002.

\begin{figure}
\centering
\includegraphics[scale=0.55]{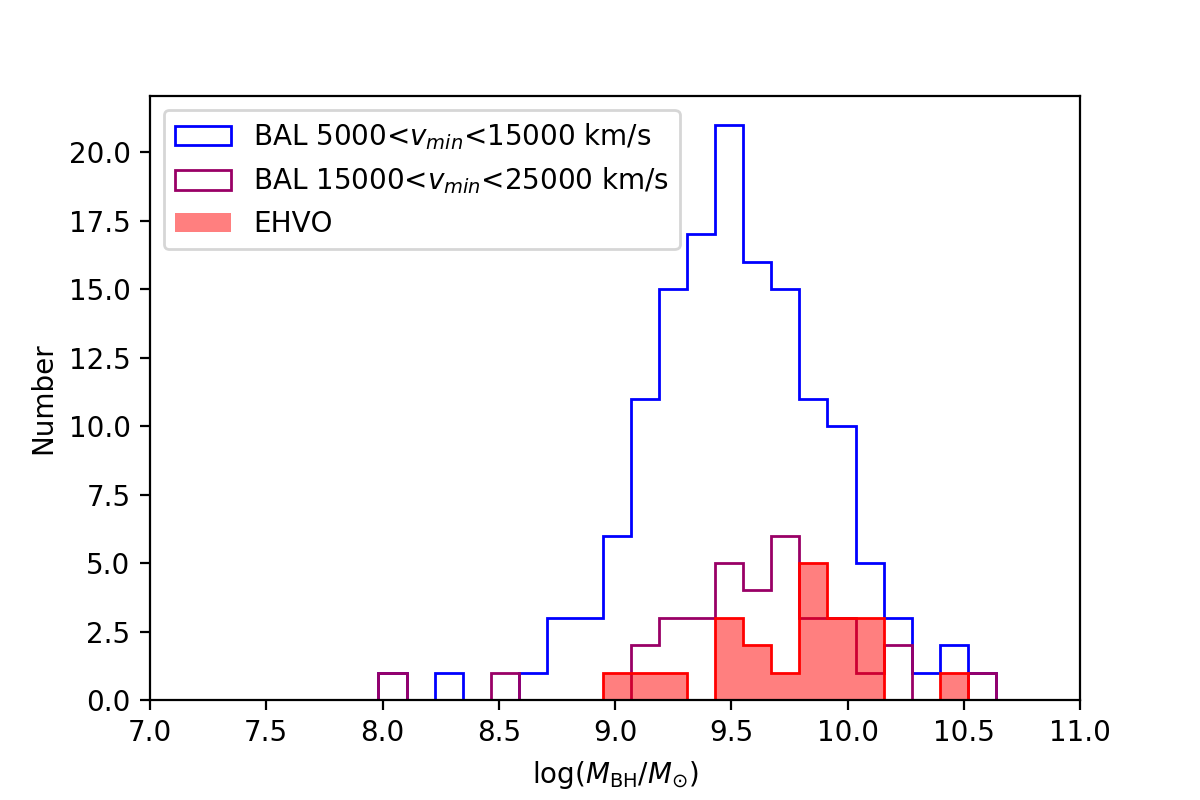}
\includegraphics[scale=0.55]{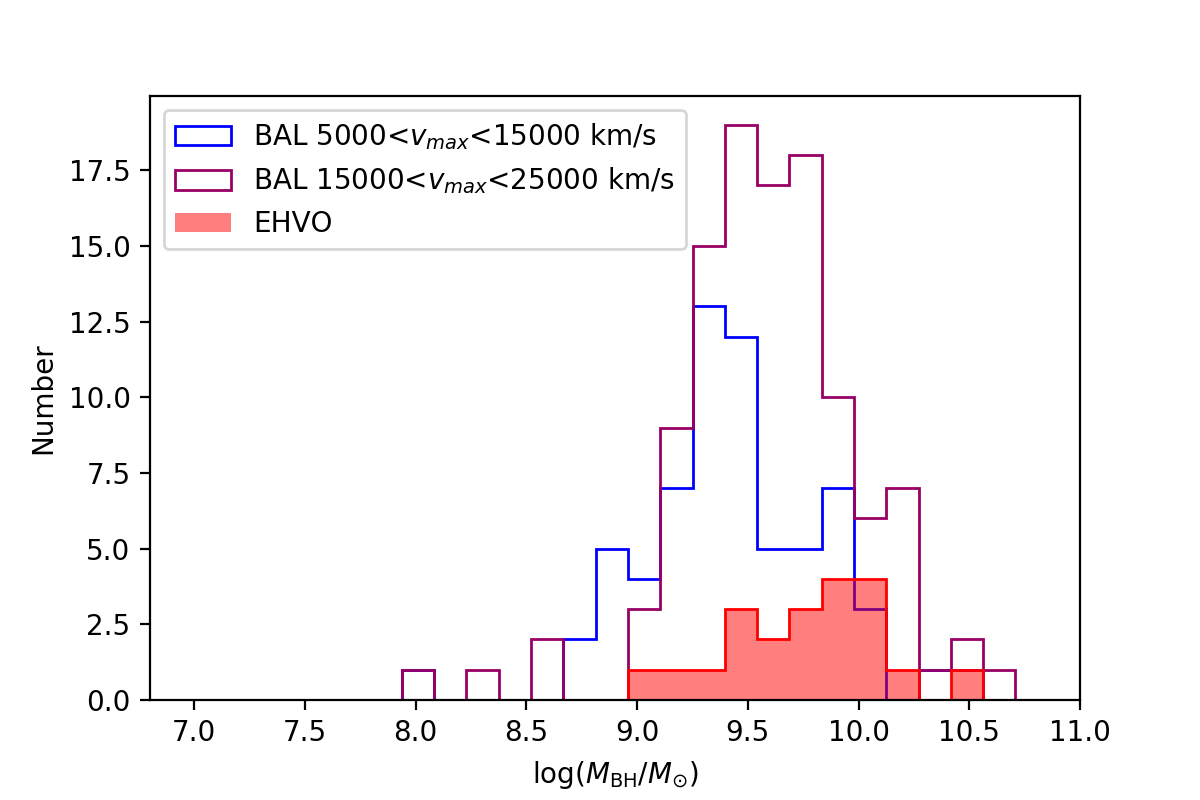}
\caption{Top: distribution of $\log(M_{BH}/M_\odot)$ values for quasars with at least one BAL with width larger than 1000 {\kms} and outflowing with 5000 $< v_{min} <$ 15,000 {\kms} (BAL in blue), quasars with at least one BAL with width larger than 1000 {\kms} and outflowing with 15,000 $< v_{min} <$ 25,000 {\kms} (BAL in burgundy), and quasars with at least one BAL with width larger than 1000 {\kms} at $v_{min} >$ 30,000 {\kms} (EHVO in red). The frequencies for EHVO quasars in each bin have not been multiplied by any factor. Bottom: similar to top figure, displaying $v_{max}$ instead of $v_{min}$. A progression toward larger values of $M_{\rm BH}$ is observed as we increase both $v_{min}$ and $v_{max}$. }
\label{MBH_BAL}
\end{figure}

We do observe progressions in the values of $M_{\rm BH}$ with increasing $v_{min}$ and $v_{max}$ in BALQSOs. In Figure \ref{MBH_BAL}, we show the distributions of $\log(M_{BH}/M_\odot)$ for EHVO and BALQSOs, where the BALQSOs have been divided into groups based on the minimum and maximum velocities of their {\CIV} outflows, similar to Figure \ref{Lbol_BAL}. The distributions of $\log(M_{BH}/M_\odot)$ at lower velocities (both for $v_{min}$ and $v_{max}$) are clearly distinct from the distribution of $\log(M_{BH}/M_\odot)$ values for quasars with larger velocities, and a progression toward larger values is observed in both cases. EHVO quasars fit correctly in this progression showing the largest values of black hole masses. 

\begin{figure}
\centering
\includegraphics[scale=0.55]{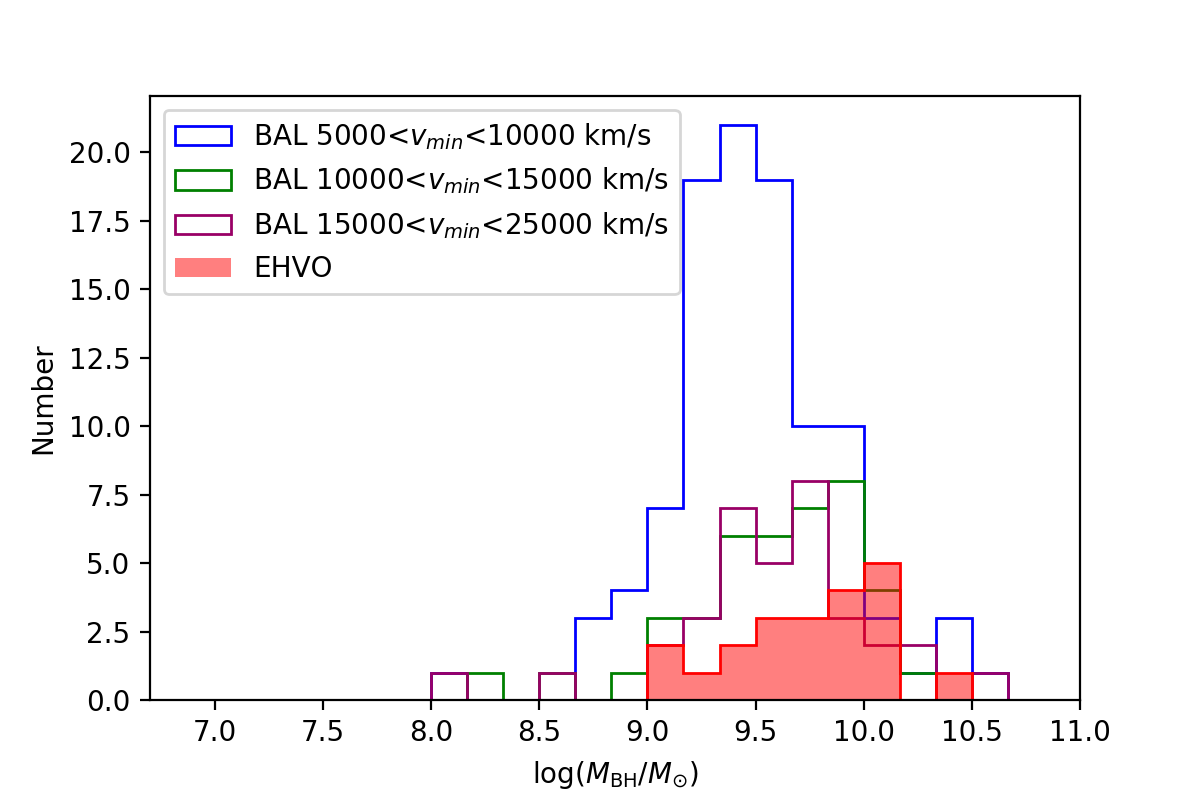}
\includegraphics[scale=0.55]{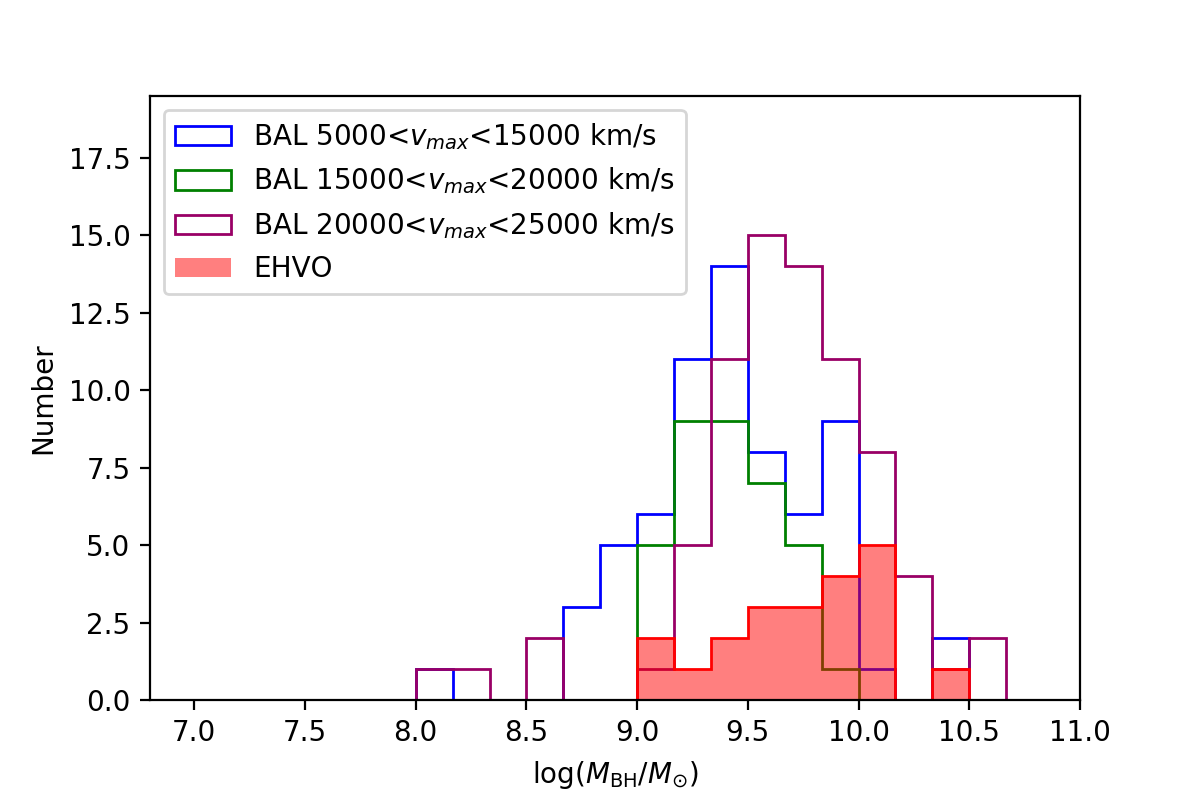}
\caption{Top: similar to Figure \ref{MBH_BAL}, but with BALQSOs divided into quasars with at least one BAL with width larger than 1000 {\kms} and outflowing with 5000 $< v_{min} <$ 10,000 {\kms} (BAL in light blue), outflowing with 10,000 $< v_{min} <$ 15,000 {\kms} (BAL in green), and outflowing with 15,000 $< v_{min} <$ 25,000 {\kms} (BAL in burgundy). Bottom: similar to the top plot, but with BALQSOs divided into quasars with at least one BAL with width larger than 1000 {\kms} and outflowing with 5000 $< v_{max} <$ 15,000 {\kms} (BAL in light blue), outflowing with 15,000 $< v_{max} <$ 20,000 {\kms} (BAL in green), and outflowing with 20,000 $< v_{max} <$ 25,000 {\kms} (BAL in burgundy). The progression toward larger $M_{\rm BH}$ as we increase the velocity values is observed in both cases: Top figure shows that the gap occurs between quasars with the smaller $v_{min}$ (5,000 $< v_{min} <$ 10,000 {\kms}) and quasars with larger $v_{min}$. Bottom figure shows the largest gap occurs between BALQSOs with the largest $v_{max}$ (20,000 $< v_{max} <$ 25,000 {\kms}) and quasars with smaller $v_{max}$. The distribution of $M_{\rm BH}$ values for EHVO quasars shows that they are similar to BALQSOs with $v_{min} >$ 10,000 {\kms} and to BALQSOs with $v_{max} >$ 20,000 {\kms}, although the median values of $M_{\rm BH}$ in EHVO quasars follow an upwards progression.}
\label{MBH_BAL_progression}
\end{figure}

To study this progression further, we subdivided the BALQSO population into smaller velocity bins. Due to small numbers we did not divide into two subgroups the BALQSOs with 15,000 $< v_{min} < $ 25,000 {\kms} and those with 5000 $< v_{max}< $ 15,000 {\kms}.
Figure \ref{MBH_BAL_progression} shows the subsets of BALQSOs that are particularly distinct in each case. The largest gaps occur between BALQSOs with 5000 $< v_{min} < $ 10,000 {\kms} and those with larger $v_{min}$: the median of $\log(M_{\rm BH}/M_\odot)$ for the latter population is 9.46, while all the subsets of BALQSOs with larger $v_{min}$ show median values of $\log(M_{\rm BH}/M_\odot) = $ 9.59$\pm$0.01. Similarly, the largest gap at $v_{max}$ occurs between BALQSOs with outflows at 20,000 $< v_{max} < $ 25,000 {\kms} and those with smaller $v_{max}$: the median of the latter is 9.63, while the other BALQSOs subsets show median values of $\log(M_{\rm BH}/M_\odot)$=$9.44\pm0.06$. 
Naturally, again, EHVO quasars fit as expected in this progression showing the largest median value of $\log(M_{\rm BH}/M_\odot)$ (9.82; see Tables \ref{Table4} and \ref{Table5}). 

In summary, the distribution of $M_{\rm BH}$ of EHVO quasars is significantly distinct from the distribution of values for BALQSOs with the lowest velocities, but it is more similar to those BALQSOs with the largest velocities. The EHVO quasar population is distinct from BALQSOs with 5000 $< v_{min} < $ 10,000 {\kms} ($p$-value of 0.005), but when comparing it to BALQSOs with larger $v_{min}$, all $p$-values are larger than 0.07. Comparing the values for EHVO quasars to the subsets of BALQSOs in $v_{max}$, all $p$-values are smaller than 0.01 except for BALQSOs with outflows with and 20,000 $< v_{max} < $ 25,000 {\kms}, for which the $p$-value is 0.13. Due to the small number of EHVO quasars in our current sample, we do not divide it into subsamples with different $v_{min}$ or $v_{max}$ to study potential trends within the velocity of the EHVO outflow, but this will be worth studying as we  increase the number of EHVO quasars with future samples. Finally, note that EHVO quasars are distinct from all BALQSOs, independently of their $v_{min}$ or $v_{max}$, in their distributions of $L_{\rm bol}$ values.

\subsection{Redshifts, $M_i[z=2]$, and Reddening of EHVO quasars, BALQSOs, and the Parent Sample}
\label{sec4.5}

\begin{figure}
\centering
\includegraphics[scale=0.55]{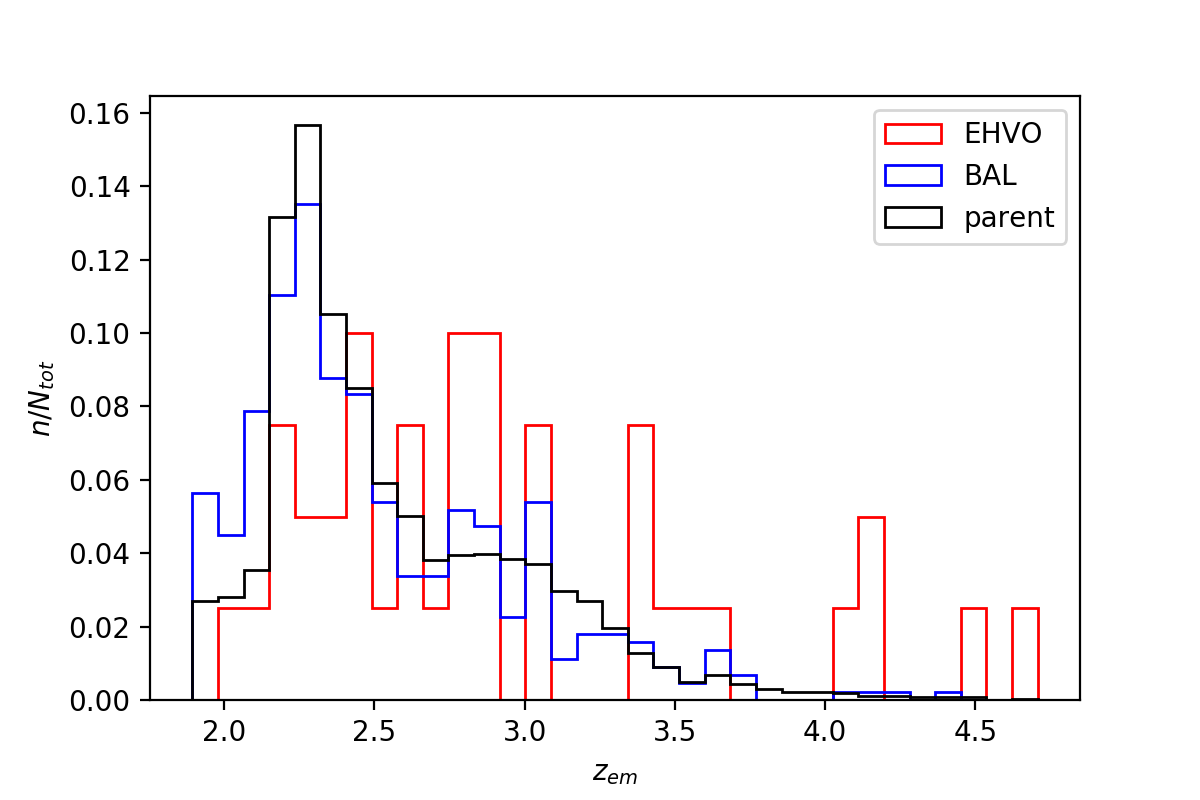}
\caption{Distribution of $z_{\rm em}$, scaled by the total number of objects in each category, for the parent sample (black) quasars, BAL quasars (blue), and EHVO quasars (red). BALQSOs and the parent sample track each other very closely, except at the lower-redshifts tail. 
While we have a smaller number of cases of EHVO quasars than in the other two categories, the current sample of 40 EHVO quasars does not seem to decrease as redshift increases as drastically as the parent and BALQSO samples do.}
\label{zemNtot}
\end{figure}

\begin{figure}
\centering
\includegraphics[scale=0.55]{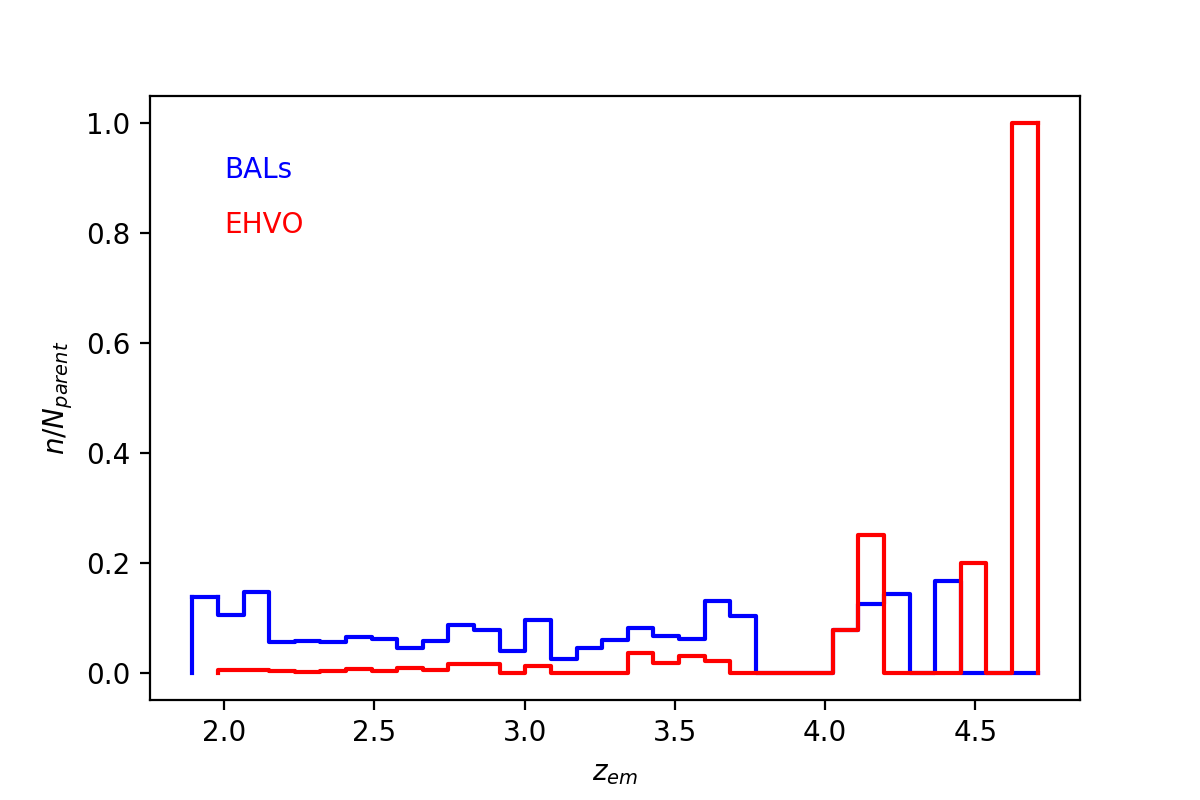}
\caption{Distribution of $z_{\rm em}$ values of BALQSOs and EHVO quasars, scaled by the total number of DR9 quasars in the parent sample in each bin, for BAL quasars (blue), and EHVO quasars (red).
While we have a smaller number of EHVO quasars than BALQSOs, they represent a similar fraction out of total quasars as redshift increases.}
\label{zemBALSEHVO2}
\end{figure}

DR9Q includes information about an array of parameters and has the advantage that we can analyze the parameter distributions for the complete samples in our study: parent sample (6743 quasars), BALQSOs (444) and EHVO quasars (40). We obtained the values for several of these parameters (such as $z_{\rm em}$, $M_i$[$z=2$], and spectral indices) in the samples, and some of the results are shown in Figures \ref{zemNtot} and \ref{zemBALSEHVO2}.

The distribution in redshifts for EHVO quasars is distinct from that of BALQSOs and the parent sample. 
In Figure \ref{zemNtot}, we show the number of quasars per redshift bin scaled by the total number of objects in each category for the three populations. 
All redshifts were taken from DR9Q unless the \citet{Hewett10} redshifts were available. EHVO quasars are more predominant at larger redshifts.
The median of the redshift value of EHVO quasars is $z_{\rm em} =$ 2.80, that of the parent sample is 2.42, and that of BALQSOs is 2.39. According to statistical tests, none of the populations are drawn from each other, as all the $p$-values are populations are smaller than \num{9e-04}. 
Obviously, the cutoff at lower redshift ($z_{\rm em}>$1.9) is artificial; we imposed it to be able to find these type of outflows in the first place in the SDSS sample.
However, beyond $z_{\rm em}>$1.9, the distribution of $z_{\rm em}$ for EHVO quasars does not follow the parent or the BALQSO distributions, both of which show a peak at $z_{\rm em}\sim$2.3. Almost a third of the EHVO quasar population has $z_{\rm em}>3.2$, while less than 10\% of the parent and BALQSO population show those values. 

Figure \ref{zemBALSEHVO2} shows the fraction of BAL and EHVO quasars relative to the parent population in each redshift bin. EHVO quasars are less frequent at lower redshifts, but represent a larger fraction than BALQSOs at higher redshifts ($z_{\rm em} > $4), where there are only 52 SDSS DR9 quasars. 
Indeed, the quasar with the largest redshift in our parent sample from DR9Q shows an EHVO outflow (1.0 value in Figure \ref{zemBALSEHVO2}). This is not surprising, as at least one of the three DR9Q quasars with redshift $z_{\rm em} > $7 is also an EHVO quasar (see Section \ref{sec5:2}).

DR9Q also includes information on the absolute magnitude in the $i$ band at $z=2$, $M_i[z=2]$ (see more details in \citealt{Paris12}). 
$M_i[z=2]$ can be used as a proxy for bolometric luminosity, and it is available for all the quasars in our EHVO sample, and parent and BAL samples. 
The results are similar to the $L_{\rm bol}$ results shown in \ref{sec:4.4.1}. 

\begin{figure}
\centering
\includegraphics[scale=0.55]{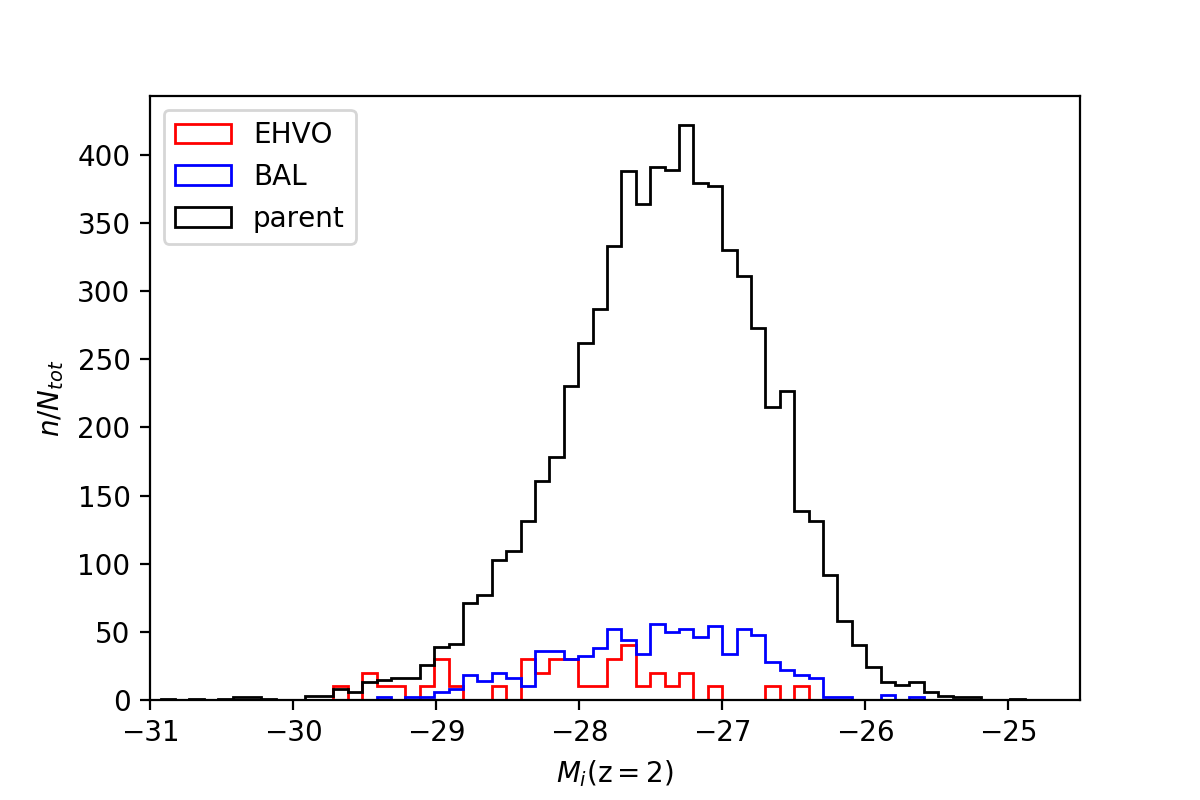}
\caption{Same as Figure \ref{Lbol}, for $M_i$[$z$=2].}
\label{Mi}
\end{figure}

Figure \ref{Mi} shows the distributions of $M_i[z=2]$ for the three studied samples, color-coded similarly to previous figures. For clarity, this figure does not include six outliers in the parent and BAL samples. As in Figure \ref{Lbol}, the population of EHVO quasars shows more negative absolute magnitudes overall (median $M_i[z=2] = -28.07$), and statistical tests show that it could not be randomly extracted from any of the other two populations ($p$-values $<$ \num{4.5e-05}). Thus, EHVO quasars show smaller absolute (negative) magnitudes and larger luminosities overall, confirming what we found from the values of $L_{\rm bol}$ in the subsample that overlaps with DR7 (see Section \ref{sec:4.4.1}).

To contrast the potential effect of reddening in the spectra of the three quasar classifications, we use the power-law slope for the continuum fit around the {\CIV} emission line provided by \citet{Shen11}, as it was the index available for all of the objects in our sample ($z_{\rm em} \geq $1.9). 
In terms of the distributions,
BAL and EHVO quasars are distinct from non-BAL quasars, with EHVO quasars tending to lie somewhere between non-BAL and BAL quasars. A two-sample K-S test rejects the null hypotheses that BAL quasars have the same spectral index distribution as non-BAL quasars with $p=10^{-32}$ (formally) and that EHVO quasars have the same spectral index distribution as non-BAL quasars with $p=0.0018$. The spectral index distributions of BAL and EHVO quasars are not different at a statistically significant level ($p=0.1037$).
Two caveats for the use of this index are (1) the wavelength region chosen is relatively narrow $[1445,1465]$ {\AA} and $[1700,1705]$ {\AA}, so it might not reflect the overall shape of the continuum, and (2) the {\FeII} emission lines redward of the {\CIV} emission were not subtracted in the spectrum prior to the continuum fitting. Therefore, we just use this spectral index as a comparison tool, but do not report numbers about individual EHVO quasars to avoid those numbers being used for the quasar's overall spectral characterization. Future work using composite spectra and including emission-line analysis will need to be carried out to define well the power-law spectral index of EHVO quasars.

\subsection{Radio Properties of EHVO Quasars, BALQSO, and the Parent Sample}
\label{sec4.6}

DR9Q also includes information about the radio loudness of the quasars, as it has been cross-correlated with the Faint Images of the Radio Sky at Twenty-Centimeters (FIRST) catalog (version 2008 July). \citet{Paris12} included the FIRST peak flux density at 20 cm ($f_\nu$(20cm), in mJy) for those quasars where FIRST had a detection within 2.$^{\prime^{\prime}}$0 of the quasar position. 

We only find one EHVO quasar (1/40, corresponding to a fraction $2.5\%^{+5.7\%}_{-2.0\%} $) where the value of $f_\nu$(20cm) was nonzero. The value is also quite small ($f_\nu$(20cm) $=$ 1.08 mJy); for reference, more than 100 quasars in the parent sample have values larger than 50 mJy. For BALQSOs, $ 3.4\%^{+1.1\%}_{-0.9\%} $ of quasars have values of $f_\nu$(20cm) larger than zero (15/444), but only one case has a value larger than 10 mJy.

Higher resolution may be required to determine the true radio nature of EHVO quasars, as radio-quiet quasars at low resolution have been found to show evidence for relativistic jets in such observations (\citealt{Jarvis19}).

\section{Discussion and Conclusions}
\label{sec:discussion}

\subsection{Summary of Results}
\label{sec5:1}

In a large parent sample (6743) of SDSS quasar spectra, with S/N $>$ 10 and $z_{\rm em} \geq$ 1.9, we have found 40 quasars with EHVOs (30,000 $< v < $ 60,000 {\kms}), observed as {\CIV} absorption in their spectra. We have investigated some of their properties as measured in the SDSS DR7 and DR9 and their relation to quasars with broad absorption at lower velocities (BALQSOs). Here we summarize the main results of this study.

\begin{itemize}

\item[1.] The 40 EHVO quasars show a large range of BI$_{\rm EHVO}$ (from 2 to 3600 {\kms}, using a minimum width of 1000 {\kms} in the definition of BI$_{\rm EHVO}$). The number of EHVO quasars includes only secure cases and it should be considered a very conservative lower limit of EHVO cases in quasar spectra (see Sections \ref{sec:search} and \ref{sec:results}).

\item[2.] Seven of these EHVO quasars are also BALQSOs as flagged in DR9Q; this number must be considered a lower limit because we reject identifying as an EHVO any absorption that could be attributed to {\SiIV} with corresponding {\CIV} at lower velocity, and that excludes many potential BALQSOs from being identified as EHVOs (see Section \ref{sec:results}).

\item[3.] We find 48 EHVO absorption features in 40 quasar spectra. EHVO absorption shows a large range of $v_{min}$, $v_{max}$, and EW values, 
but depth values all lie between 0.14 and 0.71, with most values around 0.3 (see Section \ref{sec:results}).

\item[4.] Out of the 48 absorption features, 26 have confirmed or likely corresponding {\NV} outflowing at similar speeds, and {\OVI}, at least, is likely present in 6/48 cases as well. We do not observe any confirmed Ly{$\alpha$} absorption in an EHVO (see Section \ref{sec:other_ions}).

\item[5.] Most EHVOs show no signs of {\HeII} emission in their spectra. In 20\% of cases, we find some broad emission at those wavelengths, most likely due to {\FeII} emission. Only three cases show narrow and strong emission that is likely due to {\HeII} (see Section \ref{sec4.3}).

\item[6.] We include $L_{\rm bol}$, $M_{\rm BH}$ and $L_{\rm bol}/L_{\rm Edd}$ information for EHVO quasars also found in the the SDSS DR7Q sample. The $L_{\rm bol}$ and $M_{\rm BH}$ values for the 21 resulting EHVO quasars are overall larger than for the parent sample (2883 quasars) and for BALQSOs (185). $L_{\rm bol}/L_{\rm Edd}$ values for EHVO quasars are similar to those of BALQSOs (see Sections \ref{sec4} and \ref{sec:4.4.1}).

\item[7.] We find that it is very unlikely that the distribution of $L_{\rm bol}$ and $M_{\rm BH}$ values for EHVO quasars could be obtained randomly from the parent and BALQSO samples (see Section ref{sec:4.4.1}). While little is known about the reddening and bolometric corrections of EHVO quasars, values of $M_i[z=2]$ for the whole sample support these results (see Section \ref{sec4.5}).

\item[8.] We observe a trend toward larger $M_{\rm BH}$ values as the $v_{min}$ and $v_{max}$ values of the absorption in BALQSOs increase toward EHVO values. We do not observe a trend for $L_{\rm bol}$. There are no trends with BI$_{\rm EHVO}$ or absorption depth (see Section \ref{sec:4.4.1}).

\item[9.] EHVO quasars show a large range of values of $z_{\rm em}$, but are overrepresented relative to BALQSOs at larger $z_{\rm em}$.
The median $z_{\rm em}$ for EHVO quasars is 2.80, for the parent sample it is 2.42, and for BALQSOs it is 2.39. EHVO quasars represent a large fraction of the SDSS parent quasars at redshifts larger than 4 (see Section \ref{sec4.5}).

\item[10.] The percentage of radio-loud EHVO quasars is 2.5\% (1/40). The only case with a nonzero value of $f_{\nu}$(20cm) in FIRST shows a small value  $f_{\nu}$(20cm) = 1.08 mJy) (see Section \ref{sec4.6}). Higher resolution might be needed to determine the true radio nature of EHVO quasars. 

\end{itemize}

\subsection{Are EHVO Quasars Special or Are They Quasars Observed at a Particular Orientation?}
\label{sec5:2}

The answer is probably both. In Sections \ref{sec4.5} and \ref{sec4.6}, we showed that, within the current EHVO sample (40 in DR9Q, 21 with information in DR7), EHVO quasars show different distributions from and larger median values for $L_{\rm bol}$, $M_i[z=2]$, and $M_{\rm BH}$ than both BALQSOs and the quasar parent sample overall. Thus, because they typically are found among intrinsically luminous and massive quasars, EHVO quasars appear to require particular physical properties of the central engine for their outflows to be observed predominantly at these large speeds. 

We also have found that the distribution of $M_{\rm BH}$ values shifts toward larger values as outflow speeds of BALQSOs and EHVO quasars increase. In fact, EHVO quasars and BALQSOs with the largest speeds ($v_{min} >$10,000 {\kms} and $v_{max} >$20,000 {\kms}) are not significantly distinct in their $M_{\rm BH}$ values, but they are in $L_{\rm bol}$ and $M_i[z=2]$. BALQSOs overall are on the lower half of $L_{\rm bol}$ values (Figure \ref{Lbol}). In summary, BALQSOs with the largest velocities show similar black hole masses to EHVO quasars, but the luminosities of such BALQSOs are lower than EHVOs; luminosity appears to be the parameter that distinguishes these two classes of quasars. 

It is not entirely surprising, given this, that Eddington ratios are not distinct between the three populations. Larger luminosities with similarly larger black hole masses, as EHVO quasars show, and smaller luminosities with smaller black hole masses will result in similar Eddington ratios. We find that EHVO quasars show a large range of Eddington ratios, as do BALQSOs and the quasar parent sample, with the majority of values below the Eddington limit (Figure \ref{Ledd}). 
In a recent study, \citet{Yi19} did not find significant differences in Eddington ratios for high-$z$ (3 $\leq z_{\rm em} \leq $5) BALQSOs compared to non-BALQSOs of the same luminosity and redshift. 

Our finding that EHVO quasars may show larger $L_{\rm bol}$ values overall is associated with the fact that they are predominantly found at larger $z_{\rm em}$. We find that the distribution of EHVO quasars is shifted toward larger redshifts relative to our parent DR9Q sample, which ranges from 1.9 $< z_{\rm em} <$ 4.7 (see Figure \ref{zemNtot}). EHVO quasars represent 20\%--30\% of the parent sample for certain redshift bins at $z_{\rm em}>4$. Beyond our sample, the number of known quasars decreases rapidly for $z_{\rm em} > 5$: known quasars with redshifts larger than 6 are in the hundreds (\citealt{Banados16}), and only three quasars are reported to date with $z_{\rm em} > 7$ (ULAS J1120+0641 in \citealt{Mortlock11}, ULAS J1342+0928 in \citealt{Banados18}, and DELS J0038-1527 in \citealt{Wang18}). The spectrum of DELS J0038--1527 clearly shows several outflows at extremely high velocities (see Figure 1 in \citealt{Wang18}) with $v_{max} \sim $52,000 {\kms}. The spectrum of ULAS J1342+0928 might include EHVO absorption as well, although \citet{Banados18} interpreted this absorption as the Gunn--Peterson damping wing of the Ly$\alpha$. The presence of strong EHVOs in one of the only three quasars known to date with $z_{\rm em} > 7$ is a strong indication that EHVOs might be more common at larger redshifts and a fundamental component of quasars with large $L_{\rm bol}$: DELS J0038--1527 is the most luminous quasar with $z > 7$ ($M_{1450} =-27.1$) and \citet{Wang18} estimate that DELS J0038-1527 has a bolometric luminosity of $\log (L_{bol}) =$ 47.33, similar to the median value of the EHVO quasars in our sample. The black hole mass estimate for DELS J0038--1527, log($M_{\rm BH}$)$=$ 9.12, is, however, lower than the black hole mass estimate of most DR9Q quasars. 

At low redshifts $z<1.9$, 
extensive searches for EHVO quasars have not been conducted,
and in fact, very few {\CIV} BALQSOs are known (e.g., \citealt{Ganguly07b}; \citealt{Dunn08}; \citealt{Leighly09};  \citealt{Allen11}). In Khatu et al. (2020, in preparation) we show that in a blind sample of 27 AGNs, obtained from HST Cosmic Origins Spectrograph archival data and with redshifts $z_{\rm em}< $0.2, we found no evidence of EHVOs. \citet{Hamann18} reported a potential {\CIV} outflow with speed $\sim$0.3$c$ in the spectrum of PDS 456; that was the only case we had labeled as ambiguous due to the presence of unidentified absorption blueward of the Ly$\alpha$ emission line, but there is a lack of another confirming ionic transition in the same outflow at the moment. PDS 456 shows an ultrafast outflow in its X-ray spectrum (see below). The population of EHVO quasars with $0.2<z_{\rm em}<1.9$ remains unexplored.

BALQSOs appear typically redder than non-BALQSOs, which has been attributed to being observed at lower inclinations relative to the quasar accretion disk and/or to existing in dustier environments
(\citealt{Weymann91}; \citealt{Brotherton01}; \citealt{Reichard03}). In Section \ref{sec4.5}, using \citet{Shen11} values for the power law of the fitted continuum around the {\CIV} emission line, we observe less negative overall values for BALQSOs than for the quasar parent sample, as expected.
EHVO quasars also differ statistically from the parent sample but not from BALQSOs.  
While this result is preliminary, as the power-law indexes are defined over a small wavelength region (see our discussion in\ref{sec4.5}), they might be indicative of EHVO quasars also being observed at lower inclinations and/or in dustier environments, as BALQSOs are. A future study of the values of $\alpha_{\rm UV}$ (such it was carried out in \citealt{Baskin15}) will provide better information of the potential reddening in our sources. 

EHVO quasars are not the only outflows detected with very large speeds. UFOs have been observed as Fe K-shell absorption in the X-ray spectra of nearby AGNs (predominantly Seyferts) at speeds similar or higher (0.03$c$--0.4$c$) than the EHVOs in our sample (e.g., \citealt{Chartas02}; \citealt{Tombesi10}; and references therein). These detections show predominantly narrow and weak absorption features  (EW $\leq $ 150 eV).  
Given that UFOs have been searched for mostly in local AGNs ($z_{\rm em} < $ 0.1), the central engine shows typically moderate luminosities ($\log(L_{\rm bol}) \sim$ 43--45.5 erg s$^{-1}$; this was calculated as $L_{\rm bol} = k_{\rm bol} L_{\rm ion}$ assuming a $k_{\rm bol}$ of 10,  see \citealt{Tombesi13}). If the trend between velocity and $L_{\rm bol}$ is extended to the X-ray absorbers, it might imply that UFOs in quasars with larger luminosities would show even larger velocities. The detection of UFOs in high-$z$ quasars is still rare (there are seven cases with $z_{\rm em} > $1.5 at the writing of this paper: the first one being \citealt{Chartas02}; see \citealt{Dadina18} and references therein), and mostly in lensed quasars, which make it difficult to constrain their bolometric luminosities (\citealt{Chartas14}). A systematic study of UFOs at large redshifts is prohibitively difficult at the moment. 

\vspace{1cm}

\subsection{What is Driving These Outflows? Larger Luminosities, Softer SEDs}
\label{sec5:3}

The distinctly higher luminosities of EHVO quasars vs.\ the typically lower luminosities of BALQSOs seems to favor a model where radiation pressure (e.g., \citealt{Arav94}; \citealt{Murray95}) is responsible for the largest speeds of the outflows we study in this paper. \citet{Laor02} showed that, for soft X-ray-weak quasars ($\alpha_{OX} \leq -2$), there was a correlation between the maximum velocity of the outflow ($v_{max}$) and their luminosity ($M_V$). Our results seem to initially support this finding. While we do not have direct X-ray information of the EHVO quasars, most EHVO quasars show no signs of {\HeII} $\lambda$1640.42 emission, 
and only one case shows narrow and strong {\HeII} emission (see Section \ref{sec4.3}). The lack of {\HeII} emission is an indicator of softer SED, in which the emission of higher-energy photons is weakened; BALQSOs, for example, have typically weaker {\HeII} emission (\citealt{Richards11}; \citealt{Baskin15}). In fact, \citet{Richards11} predicted that quasars with strong {\HeII} emission will be lacking absorption in their spectra; our results agree with this prediction. Surprisingly, though, the EHVO quasar with the largest BI in our sample (J164653.72+243942.2;   
BI $= 3600$ {\kms}) also shows the strongest {\HeII} emission (see Figure \ref{fig:HeIIEHVOexample}). We have already commented on this interesting case in several sections, and we will present a detailed analysis of it in Rodr\'iguez Hidalgo et al. (2020, in preparation). Except for this special case, most EHVO quasars show larger luminosities and potentially softer SEDs than is typical of other quasars.

We find that the three populations of quasars (parent, BALQSO, and EHVO) show a large range of Eddington ratios, and the three populations are indistinguishable in this regard. Large Eddington ratios are, thus, not a requirement to drive these EHVOs.

As discussed in the previous section, UFOs observed in X-ray spectra have similar speeds to EHVOs, but they are typically observed in much less luminous quasars. If there are two different scaling relationships between the outflow terminal velocity and AGN luminosity for UFOs and EHVOs, this might mean that these outflows require different launching or acceleration conditions or that they are entirely different phenomena. One of the differences already found is that UFOs are observed both in radio-quiet and radio-loud AGNs, while EHVO quasars are typically all radio-quiet at FIRST sensitivities (see Section\ref{sec4.6}). We would need to be able to detect UFOs at $z_{\rm em} \sim $ 2-5 to be able to compare them appropriately to EHVO quasars.

While the larger luminosities in EHVO quasars suggest that radiation is the driving mechanism, we cannot reject other possibilities such as magnetic driving (\citealt{deKool95}; \citealt{Proga04}; \citealt{Everett05}). Indeed, higher terminal velocities are expected in magnetically driven outflows, due to stronger centrifugal forces (\citealt{Proga07}). Comparisons to state-of-the-art magnetohydrodynamical simulations obtained with the typical luminosities and black hole masses observed in EHVO quasars will show whether these outflows can be reproduced by the models.

Given the rate of EHVO quasars within the parent population, it is not surprising that we find a lack of EHVO quasars with $\log(M_{BH}) < 9$: there might not be enough quasars per $M_{\rm BH}$ bin in DR9Q in these ranges to find secure cases of EHVO absorption. However, we do find larger proportions of EHVO quasars at large values of all $L_{\rm bol}$, $M_i[z=2]$, $M_{\rm BH}$ and $z_{\rm em}$, where the DR9Q quasars are also scarce. 
Moreover, as discussed before, we find a progression in $M_{\rm BH}$ values as the velocities of BALs increase toward EHVO values. Once we expand the sample of EHVO quasars, it will be interesting to study this progression with velocity within EHVO quasars. 
The increase of velocity with $M_{\rm BH}$ might be an indication of observing larger Keplerian velocities, indicating that the outflows are close to the AGN and the SMBH.  However, the distance between these outflows and the AGN source currently is unknown.

Overall, within the current sample, our major finding is the combination of EHVO occurring in quasars with larger luminosities and softer far-UV/X-ray SEDs (as inferred from {\HeII} emission-line strength), which potentially results in outflows easier to drive radiatively. Larger samples of EHVO quasars will allow us to study further these trends, and comparisons to theoretical simulations will help understand the physical properties that drive these outflows.

\subsection{Kinetic Power, Mass-outflow Rates, and Feedback}
\label{sec5:4}

Assuming all other conditions are equal, faster outflows are more energetic outflows. Outflows have been invoked as one of the potential regulating mechanisms to help establish the relationship found between the black hole masses in the central AGN region and the mass of the surrounding host galaxy \citep[e.g.,][]{Silk98, DiMatteo05, Springel05, Hopkins06}.
In order to determine whether these outflows could significantly affect the host galaxy around them, we need to calculate their kinetic luminosity.

The kinetic or mechanical power can be written as
\begin{equation}
    \dot{E_k} = \frac{1}{2} \dot{M}_{\rm out} (v_{\rm out})^2
\end{equation}

\noindent where the mass-outflow rate ($\dot{M}_{\rm out}$) is proportional to the velocity ($v_{\rm out}$), column density ($N_{\rm H}$), distance ($r$) and, depending on the geometry, some properties such as the global covering factor of the absorber ($f_c$). Thus, the kinetic power is largely dependent on the velocity of the outflow, as it is proportional to $(v_{\rm out})^3$. 
Comparing EHVOs to typical BAL outflows results in an increase of 1--2.5 orders of magnitude in kinetic power due solely to this velocity factor. To compare the total kinetic power, we would need good estimates of all of the other parameters ($N_{\rm H}$, $r$ and $f_c$). None of these parameters are yet constrained for EHVO quasars. However, we can set comparisons to other outflow classes by providing some rough estimates:  $N_{\rm H}$ values are likely to be smaller for EHVOs than for strong BAL outflows given that (1) the EHVO absorption does not reach large depths (see Table \ref{EHVOmeasurements}) and (2) the potential low {\SiIV}/{\CIV} ratios suggest that EHVO absorbing gas does not have extremely high optical depths. However, the difference might not be as large if the smaller depths are due to the smaller covering fraction of the absorber. 
Small covering fractions might be indicative of small distances between the AGN source and the EHVOs. 
If we assume that each of the other parameters is one order of magnitude smaller than in the case of BAL outflows, the overall kinetic power of EHVO quasars would be similar to that for BALQSOs. However, if EHVOs are at similar distances and have other similar physical and geometrical properties, their kinetic power would be much larger than that of BALQSOs. 

Larger samples of EHVO quasars will help us study trends between the outflow velocity and other quasar properties. Detailed studies of individual cases will help constrain the parameters mentioned above and, therefore, the values of mass-outflow rates and kinetic luminosity powers for EHVO quasars. This and future samples of EHVOs will become part of a database that will allow follow-up observations that will help us test physical models of acceleration mechanisms of these outflows. 

\section*{Acknowledgements} 

P.R.H. acknowledges support for herself from start-up funding provided by the UW Bothell. P.R.H. also acknowledges support from start-up funding from the College of Natural Resources and Sciences and from the Sponsored Programs Foundation at Humboldt State University through a Research, Scholarly, \& Creative Activity grant for herself, S.S.H. and C.P.Q. 
P.B.H. acknowledges support for himself, P.R.H. and V.K.  at York University from an Ontario Early Researcher Award and from the Natural Sciences and Engineering Research Council of Canada (NSERC), funding reference number 2017-05983. 
N.M. acknowledges support for A.K. from the Canadian Institute for Theoretical Astrophysics. We would like to thank the anonymous referee for their thorough review. 

Funding for SDSS-III has been provided by the Alfred P. Sloan Foundation, the Participating Institutions, the National Science Foundation, and the U.S. Department of Energy Office of Science. The SDSS-III website is http://www.sdss3.org/.
SDSS-III is managed by the Astrophysical Research Consortium for the Participating Institutions of the SDSS-III Collaboration including the University of Arizona, the Brazilian Participation Group, Brookhaven National Laboratory, Carnegie Mellon University, University of Florida, the French Participation Group, the German Participation Group, Harvard University, the Instituto de Astrofisica de Canarias, the Michigan State/Notre Dame/JINA Participation Group, Johns Hopkins University, Lawrence Berkeley National Laboratory, Max Planck Institute for Astrophysics, Max Planck Institute for Extraterrestrial Physics, New Mexico State University, New York University, Ohio State University, Pennsylvania State University, University of Portsmouth, Princeton University, the Spanish Participation Group, University of Tokyo, University of Utah, Vanderbilt University, University of Virginia, University of Washington, and Yale University. 

\bibliographystyle{aasjournal}
\bibliography{bibliography.bib}

\end{document}